\documentclass[aps]{revtex4}

\usepackage{amsmath}
\usepackage{amstext}
\usepackage{amssymb}

\usepackage{graphicx}

\def\eq#1{Eq.~(\ref{#1})}
\def\fig#1{Fig.~\ref{#1}}

\def\sec#1{Sec.~\ref{#1}}

\newcommand{\FIG}[3]{
\begin{figure}
\begin{center}
\includegraphics[width=\textwidth]{#2}
\caption{#3}
\label{#1}
\end{center}
\end{figure}}

\newcommand{\FIGTIGHT}[3]{
\begin{figure}
\begin{center}
\includegraphics[width=0.7\textwidth]{#2}
\caption{#3}
\label{#1}
\end{center}
\end{figure}}

\begin{document}

\title{Stochastic dynamics of adhesion clusters under shared constant force and with rebinding}
\author{Thorsten~Erdmann}
\author{Ulrich~S.~Schwarz}
\affiliation{Max Planck Institute of Colloids and Interfaces, 14424 Potsdam, Germany}

\begin{abstract}
Single receptor-ligand bonds have finite lifetimes, so that biological
systems can dynamically react to changes in their environment. In cell
adhesion, adhesion bonds usually act cooperatively in adhesion
clusters. Outside the cellular context, adhesion clusters can be
probed quantitatively by attaching receptors and ligands to opposing
surfaces. Here we present a detailed theoretical analysis of the
stochastic dynamics of a cluster of parallel bonds under shared
constant loading and with rebinding. Analytical solutions for the
appropriate one-step master equation are presented for special cases,
while the general case is treated with exact stochastic
simulations. If the completely dissociated state is modeled as an
absorbing boundary, mean cluster lifetime is finite and can be
calculated exactly.  We also present a detailed analysis of
fluctuation effects and discuss various approximations to the full
stochastic description.
\end{abstract}

\maketitle

\section{Introduction}

Cells in a multicellular organism adhere to each other and to the
extracellular matrix through a large variety of different
receptor-ligand bonds \cite{b:albe02}. Although not probed this way in
traditional affinity experiments, adhesion bonds in physiological
situations usually have to function under mechanical load.  For
example, cell-matrix adhesion in connective tissue is mainly provided
by focal adhesions, which are based on transmembrane receptors from
the integrin family connecting the actin cytoskeleton to the
extracellular matrix. Focal adhesions of fibroblasts, the main cell
type in connective tissue, are usually loaded by actomyosin
contractility, in particular during tissue maintenance and wound
healing. An important class of adhesion contacts in endothelial sheets
are adherens junctions, which are based on transmembrane receptors
from the cadherin family connecting the actin cytoskeletons of
different cells. Endothelial tissue often is subjected to considerable
external stress and strain, for example in lung and blood
capillaries. Leukocytes circulating with the blood flow tether to and
roll on vessel walls through transmembrane receptors from the selectin
family connecting the actin cytoskeleton to carbohydrate ligands on
the opposing surface. Here contact dissociation is accelerated due to
the shear flow pulling on the cells.  In general, there are many more
physiological conditions in which adhesion clusters are subject to
forces arising from intra- or extracellular processes, including cell
motility, development and angiogenesis.

During recent years, the behavior of different adhesion
bonds under force has been investigated extensively on the level
of single molecules by dynamic force spectroscopy
\cite{c:evan01a,c:merk01,c:weis03}. This field has been pioneered by 
AFM-experiments by the Gaub group \cite{c:flor94} and later put onto a
firm theoretical basis by Evans and Ritchie \cite{c:evan97}. Because
bond rupture can be modeled in the framework of Kramers theory as
thermally assisted escape over one or several transition state
barriers, bond strength is a dynamic quantity which depends on loading
rate. Experimentally, this prediction has been impressively confirmed
for different molecular systems
\cite{c:rief97,c:kell97,c:merk99,c:sims99}. Dynamic force spectroscopy
has been implemented with different experimental techniques, including
atomic force microscopy \cite{c:rief97}, laser optical tweezers
\cite{c:kell97} and the biomembrane force probe \cite{c:merk99,c:sims99}. 
The behavior of molecular bonds under force can also be probed in
parallel plate flow chambers. Here usually the loading process is much
faster than bond dissociation, which therefore effectively occurs
under constant load \cite{c:alon95,c:pier02}. By now, dynamic force
spectroscopy has shown that adhesion bonds feature a much more
complicated behavior under force than suggested by the traditional
affinity experiments in solution \cite{b:lauf93}. Using concepts from
the theory of stochastic dynamics
\cite{c:evan97,c:izra97,c:seif98,c:heym00,c:brau04}, a binding energy
landscape can be reconstructed from the experimental data. During
recent years, this has been accomplished for many different adhesion
receptors, including integrins \cite{c:zhan02,c:li03}, cadherins
\cite{c:baum00} and selectins \cite{c:frit98,c:evan01b}. However, while 
dynamic force spectroscopy up to now has mainly been applied to single
bonds, in physiological settings adhesion receptors usually operate
cooperatively within clusters \cite{c:bell78}. Therefore the physical
description of single adhesion bonds under force now has to be
extended to clusters of adhesion bonds under force. Clusters also open
up the possibility of rebinding of broken bonds, which is known to be
essential to achieve physiological lifetimes of adhesion clusters. For
single bonds, rebinding usually cannot be studied due to elastic
recoil of the force transducer after bond rupture \cite{c:evan01a}. In
contrast, for adhesion clusters open bonds can rebind as long as other
bonds are closed, thus keeping the spatial proximity required for
rebinding. Only if the completely dissociated state is reached,
rebinding becomes impossible and the cluster disintegrates as a whole.

Although it is clear that force leads to accelerated cluster
dissociation, it is usually not known how it is distributed over the
different closed bonds in different situations of interest. In many
cases, most prominently in rolling adhesion, only few of the different
bonds are loaded to an appreciable degree, thus dissociation occurs in
a peeling fashion \cite{c:demb88,c:hamm92,c:chan00}.  However, due to
geometrical reasons, even in this case there will be a subset of bonds
which are loaded to a similar extend. In the same vein, the loading
situation at focal adhesions can also be expected to be rather
complicated. For the case of homogeneous loading, one further has to
distinguish between loading through soft and stiff springs
\cite{c:seif00}. In the latter case, all bonds are equivalent and a
mean field description can be applied \cite{c:seif02}.  In the first
case, force is shared equally between all closed bonds and the
coupling between the different bonds in the adhesion cluster is
non-trivial. Recently, dynamic force spectroscopy has been applied to
this case for the first time \cite{c:prec02}.  Here, a vesicle
functionalized with appropriate ligands is sucked into a micropipette
and pressed onto a cell. On retraction, the vesicle is peeled off from
the outside to the inside of the contact region. However, due to
rotational symmetry around the micropipette axis, all bonds in a ring
around the periphery of the contact area share the homogeneous
loading.

The equilibrium properties of adhesion clusters has been theoretically
studied before \cite{c:bell78,c:zuck95,c:lipo96,c:weik01}, mainly in
reference to experiments on vesicle adhesion through specific
ligand-receptor pairs \cite{c:albe97,c:ches98,c:brui00}. For the
non-equilibrium dissociation of adhesion clusters under force, a
deterministic model has been introduced in a seminal paper by Bell
\cite{c:bell78}. This model has been mainly used to study more
specific problems, for example leukocyte rolling in shear flow
\cite{c:hamm87}.  Recently, the deterministic Bell-model has also been
extended to treat linear loading of a cluster of adhesion bonds, which
usually is applied in dynamic force spectroscopy
\cite{c:seif00,c:seif02}.  A stochastic version of the Bell-model has
been introduced, but studied only in the large system limit and for
specific parameter values \cite{c:coze90}. Later the stochastic model
has been treated with reliability theory in the special case of
vanishing rebinding \cite{c:tees01}. Other special cases of the
stochastic model have been treated in order to evaluate specific
experiments, for example the binding probability between ligands and
receptors on opposing surfaces as a function of contact time
\cite{c:ches98,c:zhu00}. 

In this paper, we use the stochastic version of the Bell-model to
study the case of constant shared loading in comprehensive detail. In
contrast to applications to specific experiments, we focus on generic
features of the stochastic dynamics of a cluster of parallel bonds
under shared constant loading and with rebinding. A short report on
our main results has been given before \cite{uss:erdm04a}. As shown
elsewhere, the same stochastic framework as used here for the case of
constant loading can also be used to study the case of linear loading
\cite{uss:erdm04b}.  Compared with the deterministic model, the
stochastic model has several advantages: first, only the stochastic
model allows to treat the experimental situation that rebinding
becomes impossible once the completely dissociated state has been
reached. Second, it includes fluctuations and non-linear effects,
which are important for small adhesion clusters.  Third, using the
well-developed theory on master equations, the stochastic model allows
to derive analytical results for cluster lifetime as a function of
cluster size, rebinding rate and force, which are very helpful in
evaluating adhesion experiments, including rolling adhesion
\cite{uss:schw04a}.

In the following, we consider the situation in which a certain number
of bound adhesion receptors has been clustered and connected to some
force-bearing structure. We then ask how strongly the force
accelerates dissociation, and in which sense dissociation can be
balanced by rebinding. Since we are concerned with generic features of
contact stability, our model does not consider spatial or
concentration degrees of freedom. In \sec{sec:MasterEquation}, we
define the stochastic variant of the deterministic Bell-model, which
has three dimensionless parameters. Next we introduce the appropriate
one-step master equation describing the stochastic dynamics of an
adhesion cluster under shared constant force and with rebinding. We
also explain how this master equation can be solved numerically with
the Gillespie algorithm for exact stochastic simulations. In the two
following sections, we discuss two special cases of the model in which
considerable analytical progress can be made. In each case, we first
discuss deterministic results, and then turn to the full stochastic
model. In \sec{sec:VR}, we discuss the case of vanishing rebinding. In
this case, broken bonds cannot be reformed and the number of closed
bonds in the adhesion clusters decreases in a unique sequence of
rupture events.  This can be used to construct a solution for the
master equation and to derive an expression for the average lifetime
of an adhesion clusters. In \sec{sec:VF}, we discuss the case of
vanishing force. In this case, we deal with a linear problem and
analytical solutions of the master equation can be derived for a
reflecting boundary. They can be used in turn to derive an
approximation for the case with an absorbing boundary. Cluster
lifetime can be calculated exactly as mean first passage time using
Laplace techniques. In \sec{sec:FR} we consider the general case with
finite rebinding and finite force. Although full analytical solutions
are only feasible in the case of small clusters, cluster lifetime can
be calculated exactly for arbitrary cluster size. For larger clusters,
full solutions of the master equation are obtained by exact stochastic
simulations.  Simulations are also essential to characterize single
unbinding trajecories and to understand the role of fluctuations. We
close in \sec{sec:discussion} with a discussion of experimental
issues.

\section{Master equation}
\label{sec:MasterEquation}

\subsection{Derivation}

\FIGTIGHT{fig:cartoon}%
{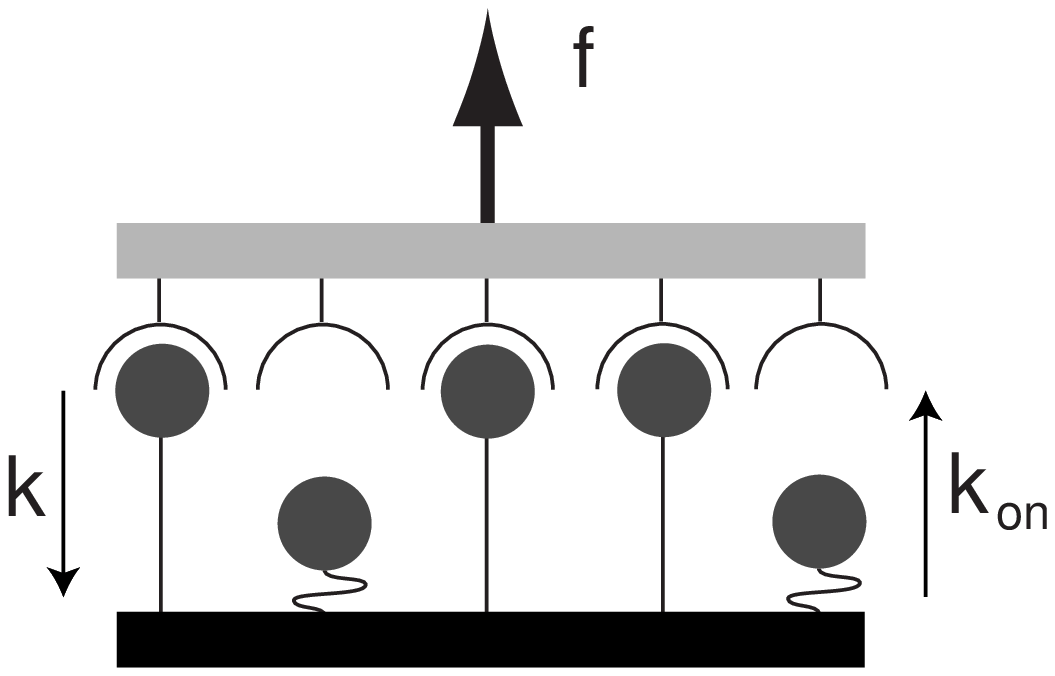}%
{Schematic representation of an adhesion cluster under constant shared
force: there are $N_t = 5$ receptor-ligand pairs, $i = 3$ of which are
closed and equally share the load $f$. A closed bond ruptures with the
dissociation rate $k = k_0 e^{f/i}$. The $N_t - i = 2$ open bonds rebind
with the force-independent association rate $k_{on}$. Our model has
three parameters: cluster size $N_t$, dimensionless rebinding rate
$\gamma = k_{on} / k_0$ and dimensionless force $f$.}

The rupture of molecular bonds can be modelled in the framework of
Kramers theory as thermally activated escape over a transition state
barrier \cite{c:evan97,c:izra97,c:seif98}. Assuming an infinitely
sharp transition state barrier leads to the so-called \emph{Bell
equation} for the single molecule dissociation rate as a function of
force, $k = k_0 e^{F / F_b}$ \cite{c:bell78}.  Here
the force scale $F_b = k_B T / x_b$ is set by thermal
energy $k_B T$ and the distance $x_b$ between the potential minimum
and the transition state barrier along the reaction coordinate of
rupture. For a typical value $x_b \simeq 1$ nm and physiological
temperature $T \simeq 300$ K, we find the typical force scale $F_b
\simeq 4$ pN. Physiological loading has indeed been found to be in the
pN-range, both for cell-matrix adhesion
\cite{uss:bala01,uss:schw02b,c:tan03} and rolling adhesion
\cite{c:alon95,c:alon97}. Values for $k_0$ and $F_b$ have been
measured during recent years with dynamic force spectroscopy for
different receptor-ligand systems, including integrins
\cite{c:zhan02,c:li03}, cadherins \cite{c:baum00} and selectins 
\cite{c:frit98,c:evan01b}. While the dissociation
rate $k$ depends mainly on the internal structure of a bond, the
association rate $k_{on}$ includes the formation of an encouter
complex and therefore depends on the details of the
situation under consideration. It is very difficult to determine
$k_{on}$ experimentally, especially in the case of cell adhesion, when
the interacting molecules are anchored to opposing surfaces
\cite{c:ches98,c:zhu00,c:orse01}. In order to focus on
the generic features of cluster stability, here we 
assume that $k_{on}$ is a force-independent constant,
in accordance with earlier theoretical work 
\cite{c:bell78,c:hamm87,c:coze90,c:seif00}.
Future modelling might refine this assumption, considering for example
the effect of ligand-receptor separation controlled by
polymeric tethers \cite{c:jepp01,c:more03,c:more04}.

For the following, it is convenient to use dimensionless
quantities. We define dimensionless time $\tau = k_0 t$, dimensionless
force $f = F/F_b$ and dimensionless rebinding rate $\gamma =
k_{on}/k_0$. The dimensionless single molecule dissociation rate is
$k/k_0 = e^{f}$. We consider a cluster with a constant number of $N_t$
bonds, which initially are all closed and then undergo rupture and
rebinding according to the appropriate rates. Since bond rupture
is a discrete process, the stochastic dynamics of the bond cluster can
be described by the one-step master equation \cite{b:kamp92}
\begin{equation} \label{MasterEquation}
\frac{dp_i}{d\tau} = r_{i+1} p_{i+1} + g_{i-1} p_{i-1} - [ r_i + g_i ] p_i\ ,
\end{equation}
where $p_i(\tau)$ is the probability that $i$ bonds are closed at time
$\tau$. Here the $r_i$ and $g_i$ are the reverse and forward rates between the
possible states $i$ ($0 \le i \le N_t$). They follow from dissociation
and association rates of single bonds as
\begin{equation} \label{Rates}
r_i = r(i) = i e^{f/i}  \quad\text{and}\quad  g_i = g(i) = \gamma (N_t - i)\ .
\end{equation}
Our model has three parameters, namely cluster size $N_t$, rebinding
rate $\gamma$ and force $f$.  Since $i \geq 0$ should be guaranteed at
any time, $r_0 = 0$ has to be set for $f > 0$, in addition to the
definitions in \eq{Rates}. Moreover, \eq{Rates} implies $g_0 > 0$,
that is, after rupture of the last closed bond new bonds are allowed
to form. This corresponds to a reflecting boundary of the master
equation at $i = 0$. As explained above, in biological and biomimetic
situations rebinding of the completely dissociated state is usually
prevented by elastic recoil of the transducer. Therefore in the
following we set $g_0 = 0$ in order to model an absorbing boundary at
$i = 0$. Because the values for $r_0$ and $g_0$ do not follow the
general form given in \eq{Rates}, the boundary at $i=0$ is an
\emph{artificial} boundary. Concerning the upper end of the set of
states at $i = N_t$, the form $g_{N_t} = 0$ represents a reflecting
boundary and guarantees $i \leq N_t$. Thus, the upper boundary is a
\emph{natural} boundary of the master equation.

A quantity of large interest is the average number of closed bonds
$N(\tau) = \langle i \rangle = \sum_{i=1}^{N_t} i p_i(\tau)$. From the
master equation \eq{MasterEquation} one can derive \cite{b:hone90}
\begin{equation} \label{FirstMoment}
\frac{dN}{d\tau} = \sum_{i=0}^{N_t} i \frac{dp_i}{d\tau}
       = - \langle r(i) \rangle + \langle g(i) \rangle\ .
\end{equation}
If $r(i)$ and $g(i)$ were both linear functions in $i$, \eq{FirstMoment}
would become an ordinary differential equation for $N$. This suggests to study 
the deterministic equation
\begin{equation} \label{DeterministicEquation}
\frac{dN}{d\tau} = - r(\langle i\rangle) + g(\langle i\rangle) 
= - N e^{f / N} + \gamma (N_t - N) 
\end{equation}
as has been done by Bell \cite{c:bell78}. Below we will see that the
analysis of this equation gives valuable insight into the generic
features of our model. However, it is important to note that for
$f > 0$, the reverse rate $r(i)$ in \eq{Rates} is non-linear in $i$
and the average in \eq{FirstMoment} cannot be taken. Instead lower
moments are related to higher moments and one arrives at a complicated
hierarchy of coupled differential equations. The solution of the
deterministic equation \eq{DeterministicEquation} will therefore
deviate from the average number of closed bonds obtained from the
solution of the master equation \eq{MasterEquation}. The same problem
arises for the higher moments. For example, for the variance $\sigma_N^2
= \langle i^2 \rangle - \langle i \rangle^2$ one can derive
\cite{b:hone90}
\begin{equation} \label{SecondMoment}
\frac{d \sigma_N^2}{dt} = 
\langle g(i) + r(i) \rangle\ + 2 \langle (i-\langle i \rangle) (g(i)-r(i)) \rangle\ ,
\end{equation}
where again the average cannot be taken. As an approximate treatment,
one can expand $r(i)$ in a Taylor series around the average for $i$,
$\langle i \rangle = N$ \cite{b:kamp92}. Restricting the expansion to
second order, thus assuming a Gaussian distribution, leads to the
following equations
\begin{align} \label{BothMoments}
\frac{dN}{d\tau} &= - N e^{f/N} + \gamma (N_t - N)
- \sigma_N^2 e^{f/N} \frac{f^2}{2 N^3}\ , \\ 
\frac{d\sigma_N^2}{d\tau} &=  N e^{f/N} + \gamma (N_t-N) 
- \sigma_N^2\left(e^{f/N}\left(2-\frac{2 f}{N}-\frac{f^2}{2 N^3}\right)+\gamma\right)\ .
\end{align}
In principle, these equations can be solved by numerical integration.
However, it is much more instructive to consider the original master
equation.  Moreover, the deterministic equation
\eq{DeterministicEquation} and its improved version \eq{BothMoments}
cannot describe the effect of an absorbing boundary at $i = 0$. In
order to consider this experimentally relevant case, one has to study
the master equation \eq{MasterEquation} with the rates given in
\eq{Rates}. Finally, only the full stochastic analysis reveals the
detailed effect of fluctuations.

\subsection{Numerical solution}
\label{sec:Master_equation_numerical}

Below we will present analytical solutions for several special cases
of the master equation. In the general case, we numerically solve the
master equation by Monte Carlo methods.  In detail, for each set of
parameter values $N_t$, $f$ and $\gamma$, we generate between $10^4$
and $10^6$ trajectories with the help of the Gillespie algorithm for
exact stochastic simulations \cite{c:gill76,c:gill77}. By averaging
for given time $\tau$ over the different simulation trajectories, we
obtain the desired probability distributions
$\{p_i(\tau)\}_{i=0}^{N_t}$. In general it is also rather instructive
to study single simulation trajectories, because their specific
features are expected to be characteristic also for experimental
trajectories.

The Gillespie algorithm was originally developed for exact simulation
of the stochastic dynamics of coupled chemical reactions.  Applied to
our case, open and closed bonds correspond to two different species of
molecules and the transition between these two species, that is
rupture and rebinding, correspond to chemical reactions.  The
Gillespie algorithm is very efficient because rather than discretizing
time in small steps, it generates jumps between subsequent
reactions. The basic quantity of the Gillespie algorithm is the
probability $P(\mu,\tau|\tau_0,X) d\tau$ that the next reaction occurs
in the time interval $[\tau_0 +\tau,
\tau_0 + \tau + d\tau]$ and is of type $\mu$ under the condition that
at time $\tau_0$ the system is in state $X$.  In our case, $\mu$ has
only two values corresponding to rupture and rebinding, and the state
$X$ of the system is completely described by the number of closed
bonds $i$. Since the rupture and rebinding rates from \eq{Rates} are
constant between subsequent events, $P$ does not depend on absolute
time $\tau_0$. In fact it reads
\begin{equation}
P(\mu,\tau|\tau_0,X) = P(\mu,\tau|i) = P_0(\tau|i) a_{\mu}
\end{equation}
where $P_0$ is the probability that no reaction occurs in the time
interval $[0,\tau]$ and $a_{\mu}$ is the reaction rate for reaction
$\mu$. $P_0$ satisfies the differential equation
\begin{equation}
\frac{dP_0}{d\tau} = - \left(\sum_{\mu} a_{\mu}\right) P_0\ ,
\end{equation}
and the initial condition $P_0(0) = 1$, therefore $P_0(\tau|i) = e^{-
\left(\sum_{\mu} a_{\mu}\right) \tau}$.  $P(\mu,\tau|i)$ is properly
normalised to unity as can be shown by integrating over time and
summing over reactions. The Gillespie algorithm generates trajectories
in which subsequent reactions are separated by the following rule. In
the absence of other reactions, the probability for a reaction
$\mu$ in the time interval $[\tau,\tau+d\tau]$ is given by
\begin{equation}\label{eq:pmu}
p_{\mu}(\tau) = a _{\mu} e^{-a_{\mu}\tau} d\tau\ .
\end{equation}
The integral  
\begin{equation}
F_{\mu}(\tau) = \int_{0}^{\tau} p_{\mu}(\tau^{\prime}) d\tau^{\prime} 
= 1 - e^{-a_{\mu}\tau}
\end{equation}
is the probability for a reaction occuring until time $\tau$. It
increases strictly monotonically from $0$ to unity and thus can be
inverted. In order to generate a random variable $\tau_{\mu}$ which is
distributed according to \eq{eq:pmu}, one generates a random number
$\xi$ which is uniformly distributed over the interval $[0,1]$ and
inserts it into the formula
\begin{equation}
\tau_{\mu} = - \frac{\ln(\xi)}{a_{\mu}}\ .
\end{equation}
This is done for each type of reaction, leading to a set of times
$\tau_{\mu}$. The time for the next reaction is then chosen as the
smallest $\tau_{\mu}$, that is $\tau = \min_{\mu}(\tau_{\mu})$.
As shown in \cite{c:gill76,c:gill77}, this rule generates trajectories
with the correct distribution of times and types of subsequent
reactions. With the forward and reverse rates \eq{Rates} for rebinding
and unbinding of molecular bonds in the cluster, the random times are
determined by the functions
\begin{equation}
\tau_{f} = - \frac{\ln(\xi)}{\gamma (N_t - i)} \quad\text{and}\quad 
\tau_{r} = - \frac{\ln(\xi)}{i e^{f/i}}\ .
\end{equation}
This algorithm is exact in the sense that the only sources of
inaccuracy lie in the choice of the random number generator and the
finite number of trajectories used to calculate probability
distribution.
  
\section{Vanishing rebinding} 
\label{sec:VR}

\subsection{Deterministic analysis}

We start our analysis with the case of vanishing rebinding, $\gamma = 0$.
Then the deterministic equation \eq{DeterministicEquation} reads
\begin{equation} \label{BellConstantLoading}
\frac{dN}{d\tau} = - N e^{f / N}\ .
\end{equation}
In principle, the total number of bonds $N_t$ is irrelevant in this
case. However, it is reintroduced through the initial condition $N(0)
= N_t$. Then \eq{BellConstantLoading} is solved implicitly by
\cite{c:bell78}
\begin{equation} \label{eq:VR_REQTN}
\tau\left(N\right) = E\left(\frac{f}{N_t}\right) - E\left(\frac{f}{N}\right)
\end{equation}
where $E(z) = \int_z^{\infty} dz' e^{-z'} / z'$ is the exponential
integral. Unfortunately, the inversion for $N(\tau)$ is not possible
in general.

In the deterministic description, cluster lifetime $T_{det}$ can be
identified with the time $\tau$ at which only one last bond
exists. Setting $N(T_{det}) = 1$ in \eq{eq:VR_REQTN} gives
\begin{equation} \label{eq:VR_Tdet}
T_{det} = E\left(\frac{f}{N_t}\right) - E\left(f\right)\ .
\end{equation}
From this result, we can extract three different scaling regimes. For
small force, $f < 1$, we use the small argument expansion of the
exponential integral, $E(z) \approx - \Gamma - \ln(z)$ (where $\Gamma
= 0.577$ is the Euler constant), and find
\begin{equation}
\label{eq:VR_Tdetsmall}
T_{det} \approx \ln N_t\ .
\end{equation}
This corresponds to the familiar case of radioactive decay, when the
differential equation $dN/d\tau = - N$ leads to exponential decay
$N = N_t e^{-\tau}$. 

For intermediate force, $1 < f < N_t$, we can rewrite \eq{eq:VR_Tdet} for
the cluster lifetime as the sum of two integrals,
\begin{equation}
\label{eq:VR_Tsplit}
T_{det} = \int_{f/N_t}^1 dz \frac{e^{-z}}{z} + \int_1^{f} dz \frac{e^{-z}}{z}\ .
\end{equation}
The second integral can be estimated to be $\int_1^{f} dz (e^{-z}/z) <
\int_1^{f} dz e^{-z} = e^{-1} - e^{-f} < e^{-1}$. For the first integral,
we can expand the integrand for small arguments, leading to
$\int_{f/N_t}^1 dz (e^{-z}/z) \approx \int_{f/N_t}^1 dz (1 - z)/z 
\approx \ln(N_t/f)$. Since $N_t > f$, the second integral can
be neglected and we have
\begin{equation}\label{eq:VR_Tmod}
T_{det} \approx \ln\left( \frac{N_t}{f} \right)\ .
\end{equation}
For the time evolution of the cluster, both exponential integrals in
\eq{eq:VR_REQTN} can be replaced by the small argument approximation
as long as $N > f$ and hence the exponential decay proceeds until the
force per bond is $f/N = 1$. Thereafter the decay will be faster than
exponential due to the destabilizing effect of force.

For large force, $f > N_t$, the second term in \eq{eq:VR_Tdet} can be
neglected and we can use the large argument approximation for the
exponential integral, $E(z) \approx e^{-z}/(1+z)$, leading to \cite{c:bell78}
\begin{equation}
\label{eq:VR_Tdetlarge}
T_{det} \approx \frac{e^{- f / N_t}}{1 + f / N_t}\quad.
\end{equation}
Therefore cluster lifetime decays faster than exponential with $f /
N_t$ in this regime. Now the small argument approximation is not
applicable to either of the two terms in \eq{eq:VR_REQTN} and the
decrease in $N$ will be faster than exponential over the whole range
of time.

In summary, the analysis of the deterministic equation
\eq{BellConstantLoading} allows to identify three scaling regimes of
small, intermediate and large force. This analysis also shows that $f
/ N_t$ is an important scaling variable, which we will therefore use
to analyse also the stochastic case.
 
\subsection{Stochastic analysis}

\FIG{fig:VR_occupancy}%
{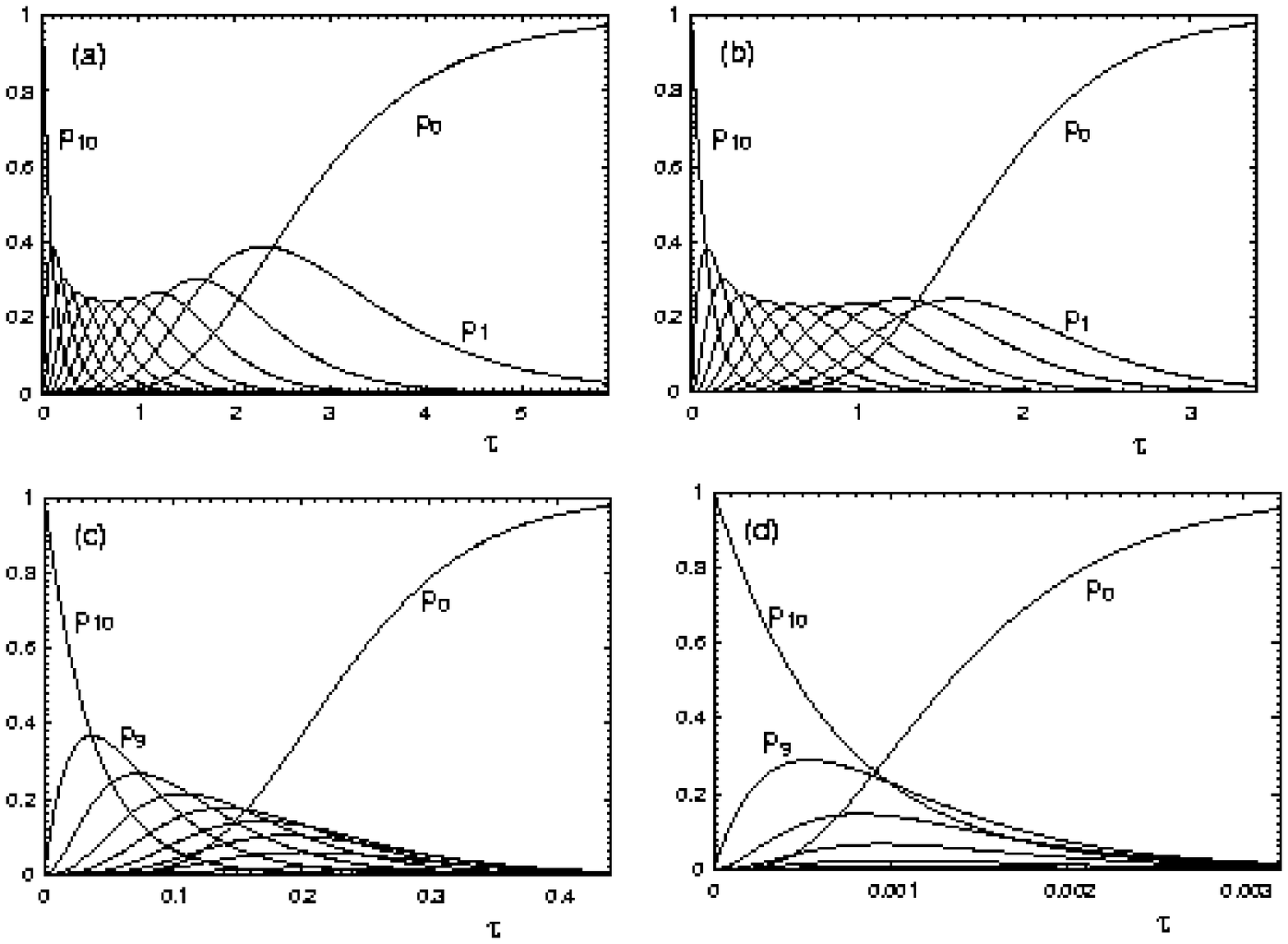}%
{The state probability $p_i(\tau)$ is the probability that $i$ bonds
are closed at time $\tau$ ($0 \le i \le N_t$). Here the $p_i$ are
plotted as a function of time $\tau$ for a cluster of initial size
$N_t = 10$ for $\gamma = 0$ and (a) $f = 0$, (b) $f = 1$, (c) $f = 10$
and (d) $f = 50$.}%

For finite force, $f > 0$, the reverse rate $r(i)$ in \eq{Rates} is
non-linear in $i$ and the boundary at $i = 0$ is artificial. Therefore
the master equation in general cannot be solved with standard
techniques. However, in the case of vanishing rebinding, $\gamma = 0$,
one can use the fact that the decay of the cluster corresponds to a
unique sequence of events, with the number of closed bonds decreasing
monotonously from $N_t$ to $0$. The transition from state $i$ (with
$i$ closed bonds present) to the state $i-1$ (with one more broken
bond) is a Poisson process with the time-independent rate $r_i$.  If
$i$ bonds are present at time $\tau$, the probability that the next bond
ruptures at time $\tau+\tau'$ is given by
\begin{equation}\label{eq:poisson_rate}
p_{i \to i-1}(\tau') = r_i e^{- r_i \tau'}\ .
\end{equation}
The state probability $p_{i-1}$ is related to the state probability
$p_{i}$ and the transition probability $p_{i \to i-1}$ by the
recursive expression
\begin{equation}\label{eq:recursivescheme}
p_{i-1}(\tau) = \int_0^\tau d\tau'\ p_{i}(\tau') p_{i \to i-1}(\tau-\tau') 
          = r_{i} e^{-r_i \tau} \int_0^\tau d\tau'\ p_{i}(\tau') e^{r_i \tau'}\ ,
\end{equation}
which uses the fact that the state $i-1$ can be reached only through the
state $i$. This scheme is solved by
\begin{equation}\label{eq:tees20}
p_i(\tau) = \left(\prod_{j=i+1}^{N_t} r(j) \right) \sum_{j=i}^{N_t}\left\{e^{-r(j)\tau}
                  \prod_{\substack{k = i\\ k \ne j}}^{N_t}\frac{1}{r(k) - r(j)} \right\}
\end{equation}
as can be shown by induction. The properties of this expression follow
from the properties of the reverse rate $r(i) = i e^{f/i}$, which for
finite force, $f > 0$, is a non-monotonous function of $i$. It
diverges for $i \to 0$, has a minimum at $i = [f]$ (the integer
closest to $f$) and grows as $i$ for $i \to \infty$.  In the unlikely
case that the value of $f$ is such that $r(j) = r(k)$ for $j \ne k$,
the limit of the expression in \eq{eq:tees20} for $r(k)\to r(j)$ has
to be taken carefully. In order to treat this case properly, one has
to replace \eq{eq:tees20} by
\begin{equation} \label{eq:tees20_a}
p_i(\tau) = \left( \prod_{j=i+1}^{N_t} r(j) \right) \sum_{j=i}^{N_t} \left\{ e^{- r(j) \tau} 
 \left(\prod_{\substack{k=i \\ r(k)\ne r(j)}}^{N_t} \frac{1}{r(k) - r(j)}\right) 
 \left(\prod_{\substack{k=i \\ r(k)=r(j) \\ k\ne j}}^{N_t} \frac{\tau}{2}\right)\right\}\ ,
\end{equation}
where we have used $\lim_{\Delta \to 0} (1 - e^{-\Delta \tau})/\Delta
= \tau$ for $\Delta = r(i) - r(k)$. In Fig.~\ref{fig:VR_occupancy} we
plot the full solution to the master equation, that is the state
probabilities $p_i(\tau)$ from \eq{eq:tees20}, for $N_t = 10$ and four
different values of force $f$. For small force, all states are
appreciably occupied during the decay, that is each of the curves is a
maximum of the set of curves during a certain period of time.  In the
long run, $p_0$ approaches unity and all other $p_i$ disappear,
because without rebinding, the cluster has to dissociate eventually.
For increasing force, the shape of the curves changes
considerably. Now the lower states (with small number of closed bonds
$i$) hardly become occupied during the decay process. For very large
force, the maximum occupancy changes directly from the initial state
$N_t$ over to the detached state $0$.

\FIG{fig:VR_Naverage}
{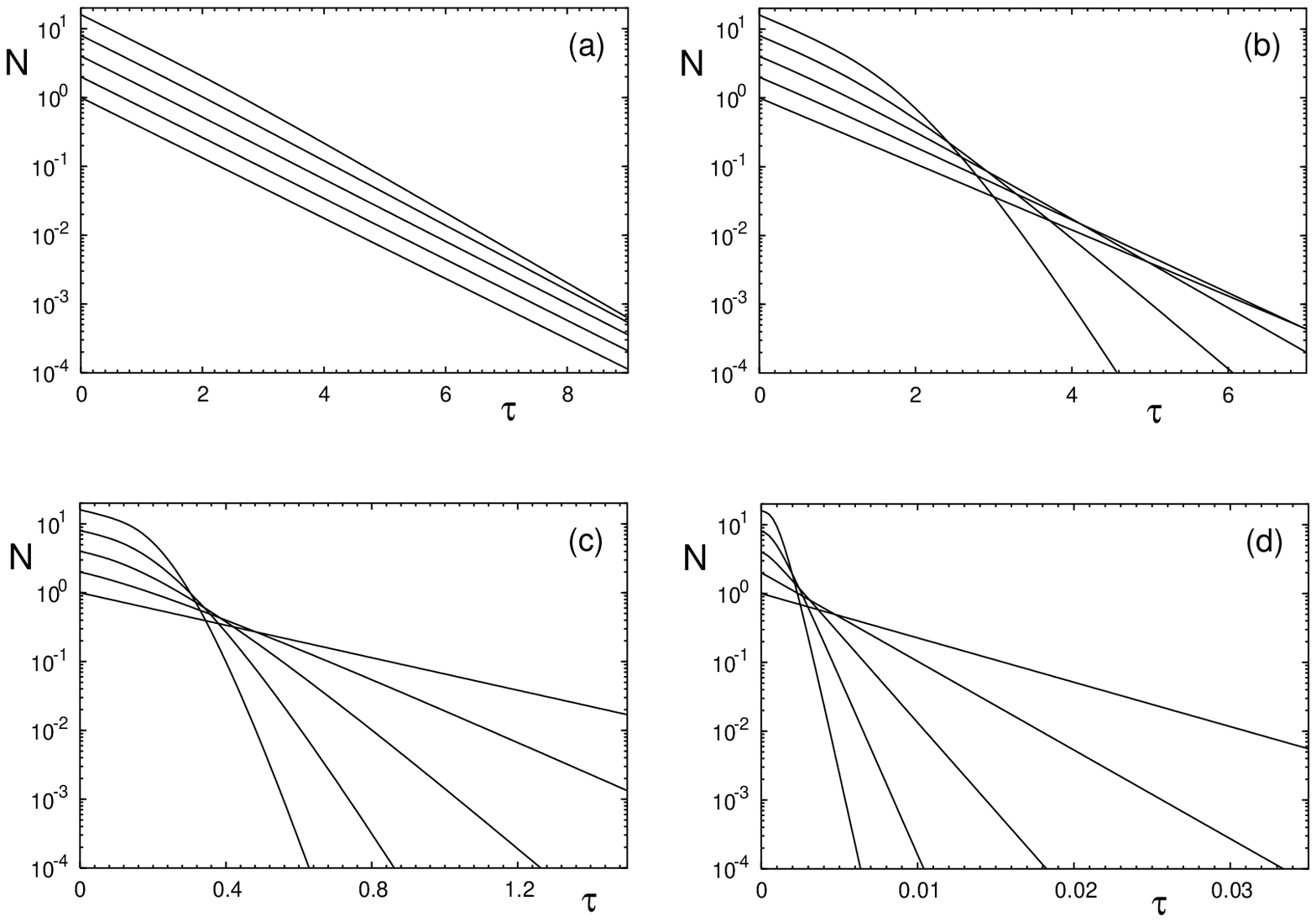}
{Average number of closed bonds $N$ as a function of time $\tau$
for different cluster sizes $N_t = 1, 2, 4, 8$ and $16$. (a) $f/N_t =
0.01$, (b) $f/N_t = 0.1$, (c) $f/N_t = 1.0$ and (d) $f/N_t = 5.0$.}

With the help of the exact solution \eq{eq:tees20}, any quantitiy of
interest can now be calculated.  One quantity of large interest is the
average number $N(\tau) = \langle i \rangle = \sum_{i=1}^{N_t} i
p_i(\tau)$ of closed bonds at time $\tau$.  For a single bond, $N$ is
simply the probability $p_1$ that the bond is attached,
\begin{equation}
N(\tau)  = p_1(\tau) = e^{-e^f \tau}\ . 
\end{equation}
For a two-bond cluster we have
\begin{equation}
N(\tau) = p_1(\tau) + 2 p_2(\tau) 
 = \frac{2}{2 - e^{f/2}} \left\{ e^{-e^f \tau} + (1 - e^{f/2}) e^{-2 e^{f/2} \tau}\right\}\ .
\end{equation}
For increasing $N_t$, the corresponding expressions become
increasingly cumbersome. In general, $N(t)$ is a sum of $N_t$
exponentials with the different relaxation rates $r(i)$ with $1 \le i
\le N_t$.  For small force, $f < 1$, $r(i) \approx i$ and the smallest
rate corresponds to $i = 1$, that is, $N \sim e^{-\tau}$ on large time
scales. In this case, clusters of any size decay with the same slope
as single bonds and the difference between single and multiple bond
rupture lies in the prefactor, not in the time-scale of average decay.
For intermediate force, $1 < f < N_t$, and large time scales, decay is
dominated by $i = [f]$, that is $N \sim e^{- f e \tau}$. Thus the
absolute value of force governs the long time behavior, with different
sizes showing up only in the prefactor. For large force, $f > N_t$,
decay at large time scales is dominated by $i = N_t$, that is $N \sim
e^{ - N_t e^{f/N_t} \tau}$.  This implies that for a given force $f$,
the largest clusters show the slowest decays in the long run.
However, if one controls $f/N_t$ rather than $f$, the cluster with the
smallest size will decay the slowest, since it is subject to the
smallest absolute force. In \fig{fig:VR_Naverage} we plot $\log N$ as
a function of time $\tau$ for different values of cluster size $N_t$
and force per initial bond, $f/N_t$. All curves initially show an
exponential decay with the rate of a single bond. For small forces
decay stays exponential for almost all times.  The larger force, the
earlier decay crosses over to the late stage regime of
super-exponential decay.

\FIG{fig:VR_Ncompare}%
{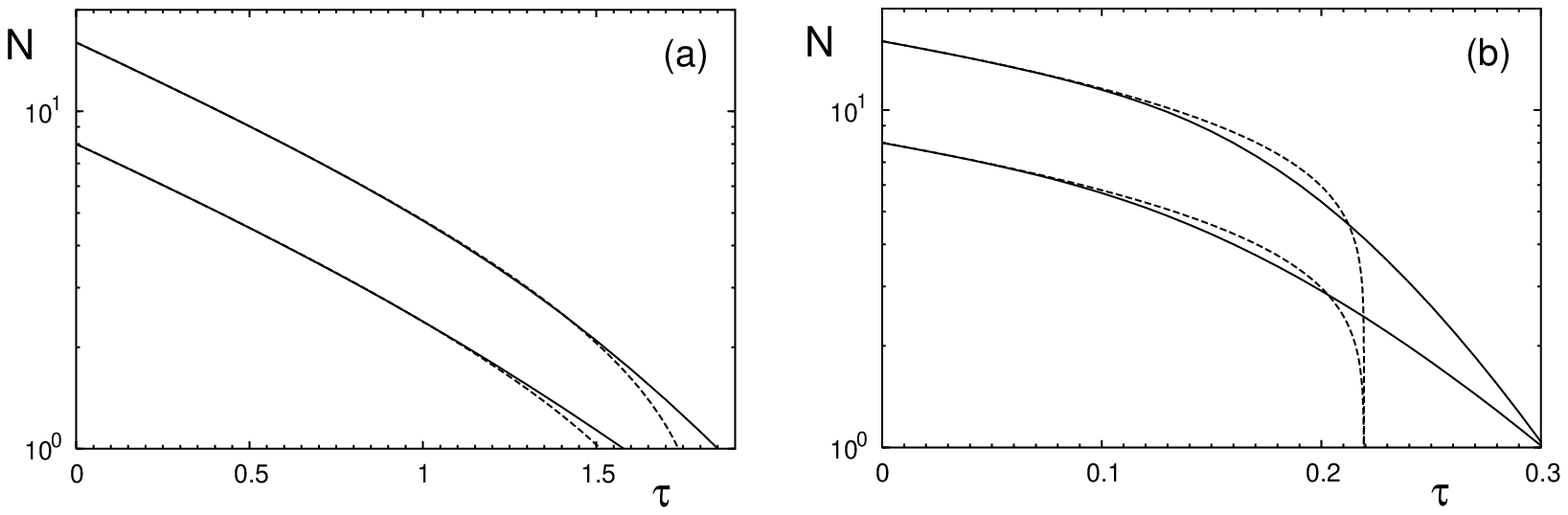}%
{Average number of closed bonds $N$ as a function of time $\tau$ for
$N_t = 8$ and $16$ and (a) $f / N_t = 0.1$ and (b) $f / N_t =
1$. Solid and dashed lines are stochastic and deterministic results,
respectively.}

As noted in \sec{sec:MasterEquation}, due to the non-linear form of
$r(i) = i e^{f/i}$ for $f > 0$, the first moment $N(\tau)$ of the
stochastic solution \eq{eq:tees20} is not identical with the function
$N(\tau)$ obtainted from the deterministic equation
\eq{BellConstantLoading}.  In \fig{fig:VR_Ncompare}, results for
$N(\tau)$ derived from the deterministic and the stochastic
description are compared to each other. For a small but non-zero
force, the non-linearity is small and the agreement between the two
results is good in the initial phase of the decay. Towards the end of
the decay strong deviations are observed.  Here, the force on each
bond grows strongly and the non-linearity of the transition rates is
large. For increasing force, fluctuations become less relevant and the
deviation between deterministic and stochastic results is increasingly
restricted to the very end of the decay process.

\FIG{fig:VR_Norbits}%
{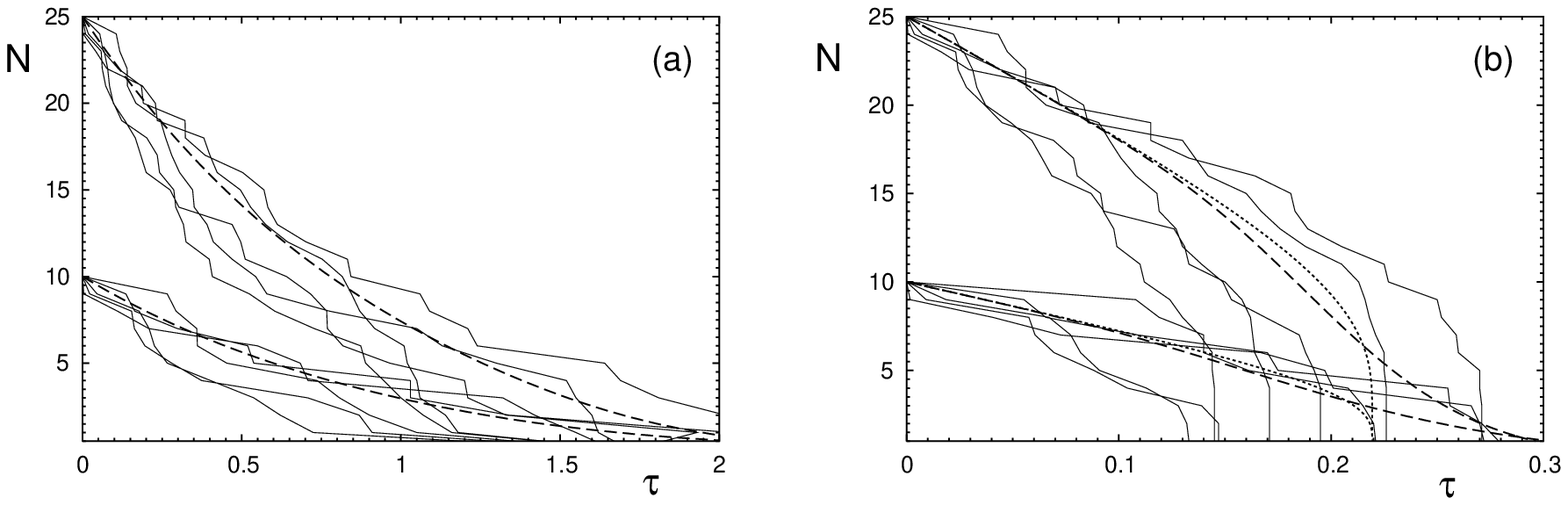}%
{Single simulation trajectories (solid lines) compared with the
average number of closed bonds $N(\tau)$ (dashed lines) for $N_t = 10$
and $25$ and (a) $f / N_t = 0.1$ and (b) $f / N_t = 1$. In (b), in
addition the deterministic results are plotted as dotted lines.}

In \fig{fig:VR_Norbits}a and b the result for the mean number of
closed bonds $N(\tau)$ is compared to single simulation trajectories
for small and large forces, respectively. The single simulation
trajectories are expected to resemble experimental realizations for
the time evolution of the number of closed bonds.  The figure shows
that for small force, the trajectories decay in a similar way as does
the average. For large force, the trajectories decay in a more abrupt
way than the averages, that is they appear to run along the average
for most of the time, but then decay rather abruptly towards the
completely dissociated state. In this case, fluctuations do not so
much affect the typical shape of the rupture trajectory, but rather
the timepoint of rupture. The reason for this typical behavior is that
a large fluctuation towards the absorbing boundary inevitably leads to
a runaway process, since force is increasingly focused on less and
less bonds due to shared loading. This type of rupture process is
similar to avalanches or cascading failures in highly connected
systems. Although rupture is rather abrupt, its timepoint is widely
distributed, leading to the smooth decrease of $N(\tau)$ observed in
the average. In the large force case in \fig{fig:VR_Norbits}b, we also
show the deterministic results for $N(\tau)$ (for the small force case
in \fig{fig:VR_Norbits}a, they hardly differ from the stochastic
results). These curves show that the abrupt decay of single simulation
trajectories at large force is somehow predicted by the deterministic
description, compare \fig{fig:VR_Ncompare}. This had to be expected
because the deterministic equation describes a representative yet
single trajectory.

\FIG{fig:VR_Tdissstribution}
{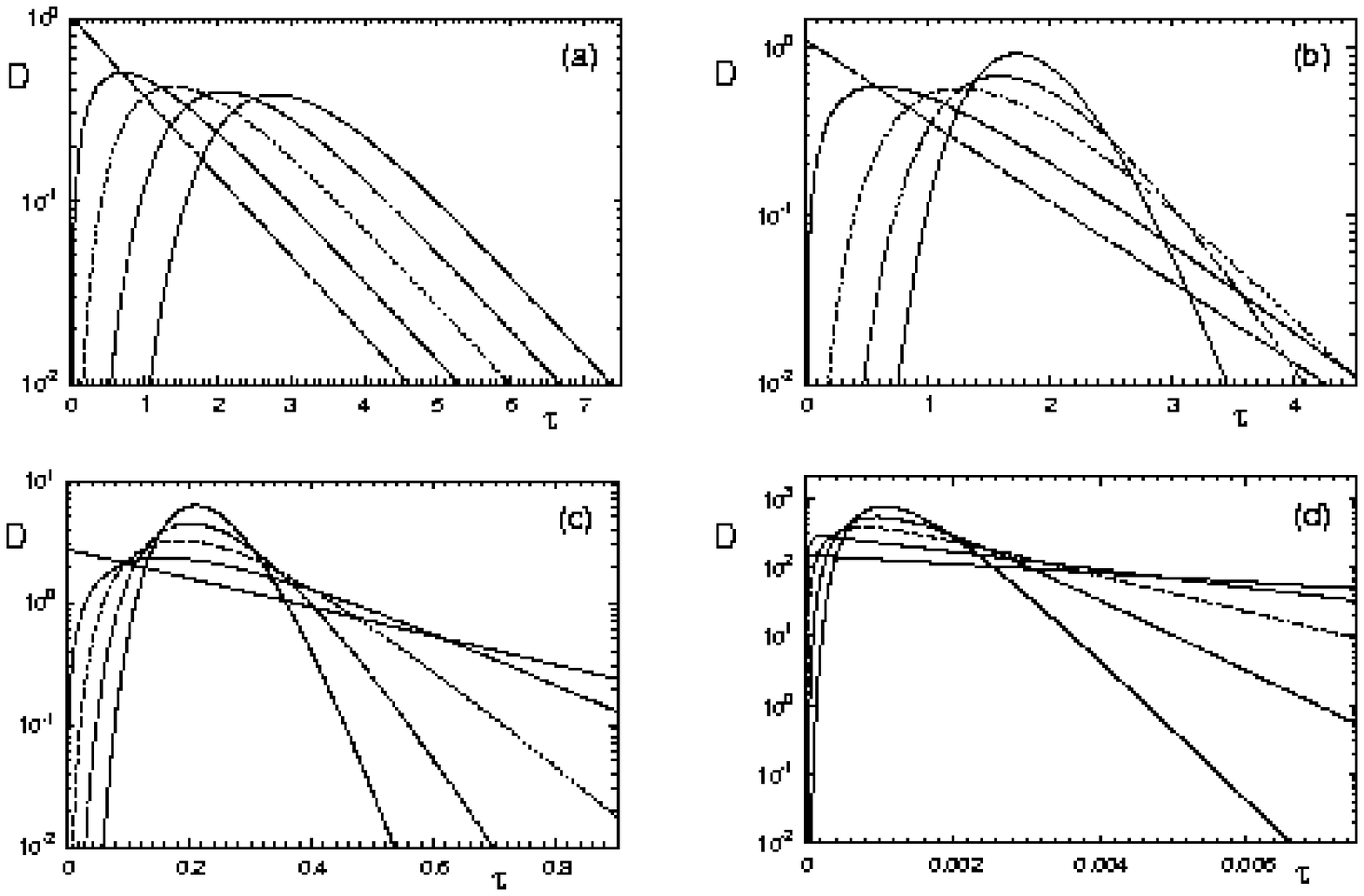}%
{Probability for dissociation of the whole cluster $D$ as a function
of time $\tau$ for $N_t = 1, 2, 4, 8$ and $16$ for (a) $f/N_t = 0.0$,
(b) $0.1$, (c) $1.0$ and (d) $5.0$.}%

The probability for dissociation of the overall cluster (that is
for rupture of the last bond) is defined by $D(\tau) = \dot p_0(\tau) =
r_1 p_1(\tau)$ and follows from \eq{eq:tees20} with the reverse rate
$r_1$ from \eq{Rates}. The resulting formula has been given before in
Ref.~\cite{c:tees01}. For a single bond it is simply $D(\tau) = e^f
e^{- e^f\tau}$. For $N_t = 2$ we have
\begin{equation}
D(\tau) = \frac{2 e^f}{2-e^{f/2}}\left(e^{-e^f\tau}-e^{-2e^{f/2}\tau}\right)\ .
\end{equation}
In the special case $f = 2 \ln 2$, the two rates $r(1)$ and $r(2)$ are
equal and we have
\begin{equation}
D(\tau)= 16 \tau e^{-4 \tau}\ .
\end{equation}
In general, as for $N(\tau)$, $D(\tau)$ is a sum of exponentials
$e^{-r(i) \tau}$ and the decrease on long time scales is governed by
the exponential which decreases the slowest. In
\fig{fig:VR_Tdissstribution} we plot $D(\tau)$ for different values of
$N_t$ and $f/N_t$ (by controlling $f/N_t$ rather than $f$, the curves
have comparable averages). The case $N_t = 1$ is a Poisson process
with simple exponential decay. For $N_t > 1$, $D(0) = 0$, because
instantaneous rupture of all bonds at $\tau = 0$ is a higher order
process.  For large times, all curves decay exponentially. For
vanishing force, $f = 0$, the curves are very similar, with the same
slope at large times. The maxima of the cluster dissociation rates for
$f = 0$ are described by $T_{max} = \ln N_t$, in agreement with the
result \eq{eq:VR_Tdetsmall} from the deterministic description. For
small $f/N_t$, the distributions are Gauss-like with small asymmetry
and variance.  For large $f/N_t$, they became Poisson-like, that is
they develop a strong asymmetry with a maximum close to zero and a
pronounced long-time tail. The reason is that in this case, decay is
dominated by rupture of the first bond, that is we are effectively
back to a single bond system (except that $D(0) = 0$ as always for
multiple bonds).

\FIGTIGHT{fig:VR_lifetime}%
{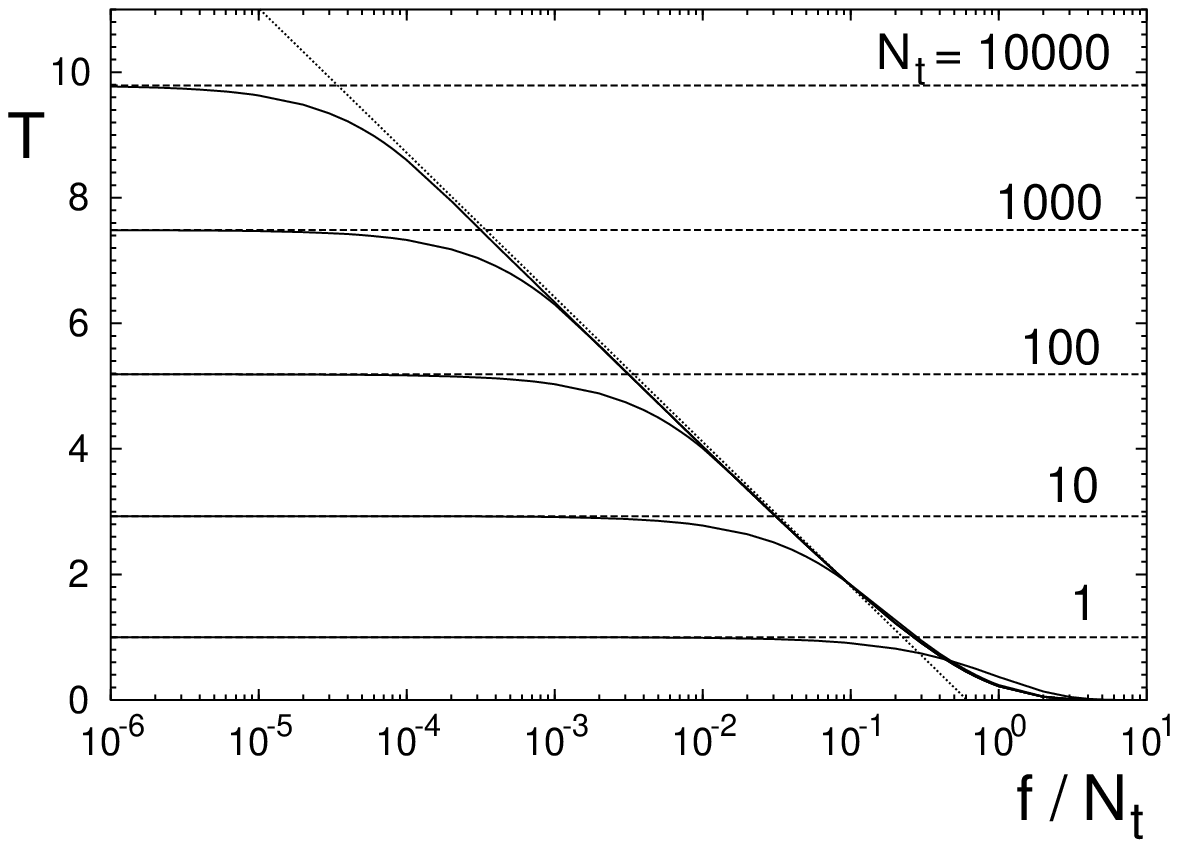}%
{Average adhesion cluster lifetime $T$ as a function of $f / N_t$ for
cluster sizes $N_t = 1, 10, 100, 1000$ and $10000$. The dashed
horizontal curves are the harmonic numbers, which are good
approximations in the small force regime, $f < 1$. The dotted curve is
the approximation $T = \ln\left( 0.61 N_t/f \right)$ for the
intermediate force regime, $1 < f < N_t$.}%

The average cluster lifetime can in principle be calculated as the
first moment of the overall dissociation rate
\begin{equation}
T = \int_0^{\infty} d\tau\ \tau\ D(\tau)\ .
\end{equation}
In practice, it has a simple form which can be derived without using
the probability distribution \eq{eq:tees20}. The waiting time spent in
state $i$ before the transition into state $i-1$ is a stochastic
variable characterised by the distribution function
\eq{eq:poisson_rate}. Its average is given by the inverse transition
rate $1/r(i)$. Since the decay process is a sequence of such
independent Poisson processes, we simply have
\begin{equation} \label{eq:lifetime}
T = \sum_{i = 1}^{N_t} \frac{1}{r(i)}\ .
\end{equation}
For $f = 0$ we get \cite{c:gold96,c:tees01}
\begin{equation} \label{eq:harmonic_number}
T = \sum_{i = 1}^{N_t} \frac{1}{i} = H_{N_t} 
\end{equation}
which are the harmonic numbers. The lifetime of a two-bond cluster is
increased by a factor $3/2 = 1.5$ with respect to the single bond, that of
the three-bond cluster by $11/6 = 1.8$, and so on. For large $N_t$
one can write \cite{c:gold96}
\begin{equation} \label{eq:lifetime_harmonic}
T \approx \ln N_t + \frac{1}{2 N_t} + \Gamma
\end{equation}
where $\Gamma = 0.577$ is the Euler constant. In fact this approximation
is very good already for small values of $N_t$. The weak (logarithmic)
dependence for large $N_t$ means that for large adhesion clusters,
size matters little since the bonds decay independently of each other
and on the same time-scale. This result differs from the deterministic
one for small force, \eq{eq:VR_Tdetsmall}, by the constant $\Gamma$
and the additional contribution $1/2N_t$, which vanishes for large
clusters. For small force, $f < 1$, \eq{eq:lifetime_harmonic} is a good
approximation for cluster lifetime $T$. For intermediate force, $1 < f
< N_t$, the reverse rate grows rapidly for states with $i < f$,
whereas for states with $i > f$, $r(i)$ remains close to
$i$. Therefore we can approximate the average lifetime of a cluster as
$H_{N_t} - H_{f}$. Using \eq{eq:lifetime_harmonic}, we get
\begin{equation} \label{eq:lifetime_mod}
T \simeq \ln \left( N_t / f \right)\ .
\end{equation}
Thus cluster size $N_t$ is now replaced by an effective size $N_t /
f$, as we have already found in the deterministic framework, compare
\eq{eq:VR_Tmod}. For large force, $f > N_t$, the only term which
contributes to \eq{eq:lifetime} is the one for the rupture of the
first bond. Then
\begin{equation}\label{eq:lifetime_large}
T \approx \frac{e^{-f/N_t}}{N_t}
\end{equation}
and we deal essentially with a single bond effect: if the first bond
breaks, all remaining bonds follow within no time (`domino effect').
This effect is also evident from the dissociation rate $D(\tau)$,
which for very large force approaches a Poisson distribution, compare
\fig{fig:VR_Tdissstribution}d.  In \eq{eq:lifetime_large}, the
numerator represents the probability for single bond rupture under
force, while the denominator represents the probability that any one
out of $N_t$ identical bonds breaks first.  Since $f > N_t$ in this
regime, the first effect dominates and $T$ increases with $N_t$. For a
given $f / N_t$, on the other hand, the lifetime decreases with
increasing $N_t$, due to the increase in absolute force. In contrast
to the deterministic result, \eq{eq:VR_Tdetlarge}, the stochastic
result \eq{eq:lifetime_large} does not scale with $f/N_t$. In
\fig{fig:VR_lifetime} we plot the average cluster lifetime $T$ 
from \eq{eq:lifetime} as a function of $f/N_t$ for different values of
$N_t$.  For small force, $f < 1$, $T$ plateaus at the value given by
the harmonic number $H_{N_t}$ according to \eq{eq:harmonic_number}. In
the regime of intermediate force, $1 < f < N_t$, all curves fall on
the master curve $T = \ln\left( 0.61 (N_t/f)\right)$, as predicted by
\eq{eq:lifetime_mod}. For large force, $f > N_t$, the scaling with $f
/ N_t$ is lost, as predicted by \eq{eq:lifetime_large}. Although
deterministic and stochastic predictions for cluster lifetime $T$ have
similar overall features, the deterministic result underestimates the
plateau at small force and predicts an incorrect scaling with $f/N_t$
at large force.

\FIG{fig:VR_cumulants}%
{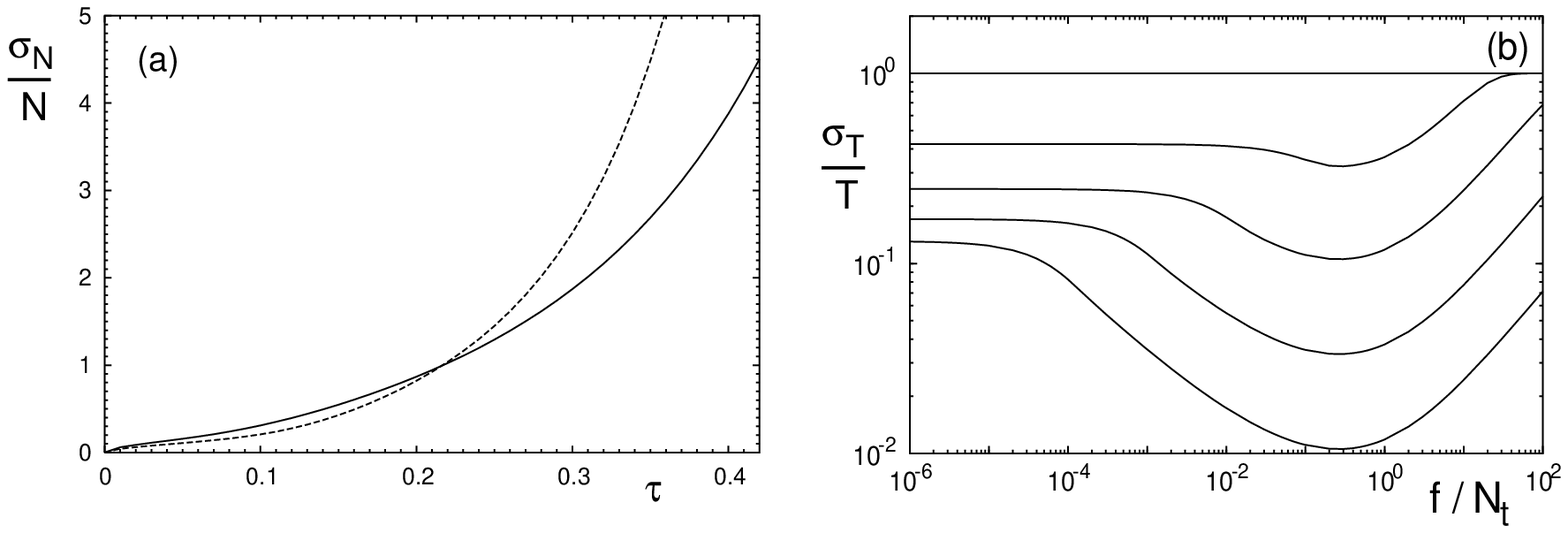}%
{(a) Variance $\sigma_N$ for the number of closed bonds in relation to
the average number of closed bonds $N$ as function of time $\tau$ for
$f/N_t = 1.0$ and $N_t = 8$ (solid) and $16$ (dashed).  (b) Variance
$\sigma_T$ for cluster lifetime in relation to average cluster
lifetime $T$ as a function of $f/N_t$ for $N_t = 1, 10, 100, 1000$ and
$10000$.}

Higher cumulants of the various distributions provide information
about the effect of fluctuations. For the number of closed bonds at
time $\tau$, the width of the distribution is described by the
variance, defined by $\sigma^2_N(\tau) = \langle i^2 \rangle - \langle
i \rangle^2$. In \fig{fig:VR_cumulants}a, we plot the relative
standard deviation, $\sigma_N(\tau) / N(\tau)$, for cluster sizes $N_t
= 8$ and $N_t = 16$. It it zero initially due to the initial condition
and diverges for large times. In regard to the distribution of cluster
lifetime, the variance $\sigma_T$ can be calculated in the same way as
the average lifetime, because for a sequence of independent stochastic
processes, all cumulants simply add up. The variance of the Poisson
process \eq{eq:poisson_rate} is $1/r^2(i)$. Therefore the variance for
cluster lifetime is
\begin{equation} \label{eq:lifetime_var}
\sigma^2_T = \sum_{i = 1}^{N_t} \frac{1}{r^2(i)}\ .
\end{equation}
For vanishing force this expression reads
\begin{equation} \label{eq:lifetime_var_zero}
\sigma^2_T = \sum_{i = 1}^{N_t} \frac{1}{i^2} 
    = \zeta(2) - \psi^{(1)}(N_t + 1)\ ,
\end{equation}
where $\psi^{(1)}(N_t + 1)$ is the trigamma function and $\zeta$ the
Riemannian $\zeta$-function. For increasing $N_t$, the variance
converges to a finite value. For zero force this limit is given by
\begin{equation}
\sigma^2_T = \sum_{i = 1}^{\infty} \frac{1}{i^2}
    = \zeta(2) = \frac{\pi^2}{6}\ ,
\end{equation}
because the trigamma function vanishes in this limit. This result is
an upper limit for the variance in general, because the reverse rate
increases with increasing force, $r(i) \geq i$. The relative standard
deviation $\sigma_T/T$ of cluster dissociation is always smaller than
unity, since $(\sum x)^2 > \sum x^2$. For single bond rupture, we have
a single Poisson process and it becomes exactly unity.  For vanishing
force and large clusters, it scales as $\sim 1 / \ln N_t$. Although it
decreases with increasing $N_T$, it does so in a different way than
the Gauss process, which decreases as $\sim 1/ N_t^{1/2}$. The reason
is that the contributions from the different subprocesses are not
constant, but decrease as rupture proceeds. For large forces, $f >
N_t$, cluster dissociation becomes a Poisson process governed by the
rupture of the first bond. Then the first term dominates in
\eq{eq:lifetime} and \eq{eq:lifetime_var}. Therefore the relative
standard deviation $\sigma_T/T \approx 1$ again. Moreover, now
$\sigma_T/T \sim 1/ N_t^{1/2}$, because now only the first $N_t - f$
subprocesses contribute, with roughly similar values, like in a
Gauss-distribution. In \fig{fig:VR_cumulants}b, we plot $\sigma_T/T$
as a function of $f/N_t$ as it crosses over between the cases of
vanishing and very large force, with a minimum around $f / N_t \approx
0.3$, that is in the intermediate force range. The narrow distribution
at intermediate force is also evident in \fig{fig:VR_Tdissstribution}.
\fig{fig:VR_cumulants}b also shows how the relative standard deviation
decreases with increasing cluster size $N_t$.  In general, the
agreement between deterministic and stochastic descriptions is best
for large cluster size $N_t$ and intermediate force $1 < f <
N_t$. However, it should also be noted that the definition of
deterministic lifetime is somehow arbitrary, because a discrete cutoff
has to be introduced in a continuum description. Especially for small
clusters the choice of the cluster size at which dissociation occurs
will have a large influence on $T$.

\section{Vanishing force}
\label{sec:VF}

\subsection{Deterministic analysis}

We now turn to the case of vanishing force, $f = 0$. Then the
deterministic equation \eq{DeterministicEquation} reads
\begin{equation} 
\frac{dN}{d\tau} = - N + \gamma (N_t - N)\ .
\end{equation}
For the initial condition $N(0) = N_t$, its solution is
\begin{equation} \label{Neq}
N(\tau) = \frac{\gamma + e^{- (1 + \gamma) \tau}}{1 + \gamma} N_t 
        = \left[ 1 + \frac{1}{\gamma} e^{- (1 + \gamma) \tau} \right] N_{eq}\ .
\end{equation}
Thus there is an exponentially fast relaxation from $N_t$ to the
equilibrium state with $N_{eq} = \gamma N_t / (1 + \gamma)$ closed
bonds. $N_{eq}$ increases linearly with the rebinding constant
$\gamma$ from $N_{eq} = 0$ for $\gamma = 0$ and saturates at $N_t$ for
$\gamma > 1$. In the deterministic description, the lifetime of the
cluster is infinite, because the completely dissociated state $N = 0$
is never reached.

\subsection{Stochastic analysis}

In the case $f = 0$, the reverse rates defined in \eq{Rates} are
linear in $i$ and $r(0)=0$ at $i=0$. Natural boundary conditions imply
$g(0) = \gamma N_t$, that is a reflecting boundary condition at $i =
0$. A linear system with natural boundary conditions can be solved
with standard techniques. For the initial condition $p_i(0) =
\delta_{i,N_t}$, a generating function has been derived by Mc Quarrie \cite{r:mcqu63}:
\begin{equation}
\label{GeneratingFunction}
G(s,\tau) = \sum_{i = 0}^{N_t} s^i p_i(\tau) = \left(\frac{(s-1)
e^{-(1+\gamma)\tau} + 1 +\gamma s}{(1+\gamma)}\right)^{N_t}\ .
\end{equation}
The state probabilities follow from the generating function as
\begin{equation} \label{eq:VF_states}
p_i(\tau) = \frac{1}{i!}\left. \frac{\partial^i G(s,\tau)}{\partial s^i} \right|_{s=0}
          = \binom{N_t}{i}\frac{\left( \gamma + e^{-(1+\gamma)\tau} \right)^i
            \left(1 - e^{-(1+\gamma)\tau}\right)^{N_t - i}}{(1+\gamma)^{N_t}}\ .
\end{equation}
One can easily check that by setting $\gamma = 0$ in
\eq{eq:VF_states}, one obtains the same result as by setting $f = 0$
in \eq{eq:tees20}. \eq{eq:VF_states} shows that the systems relaxes to
the stationary state on a dimensionless time scale $1/(1+\gamma)$,
thus the larger rebinding, the faster the system equilibrates. In the
stationary state, the state probabilities follow a binomial
distribution
\begin{equation} \label{eq:binomial}
p_i(\infty) = \binom{N_t}{i} \frac{\gamma^i}{(1+\gamma)^{N_t}}
\end{equation}
because the bonds are independent and each bond is closed and open 
with probabilities $\gamma/(1+\gamma)$ and $1/(1+\gamma)$, respectively.

The generating function also allows to calculate all moments of the
distribution:
\begin{equation}
\langle i^n \rangle = \left.\frac{\partial^nG(s,\tau)}{\partial (\ln s)^n}\right|_{s=1}\quad .
\end{equation}
Since now $r(i)$ is linear in $i$, the first moment $N(\tau) = \langle i \rangle$
is identical to the solution \eq{Neq} of the deterministic
equation. In order to assess the role of fluctuations, we calculate
the variance:
\begin{equation}\label{eq:VF_Nvariance} 
\sigma^2_N(\tau) = \langle i^2 \rangle - \langle i \rangle^2 
= \frac{(1 - e^{-(1 +\gamma)\tau})}{(1+\gamma)}N(\tau)\ .  
\end{equation} 
The relative standard deviation $\sigma_N/N$ essentially scales as
$N^{-1/2}$ for all times, thus fluctuation effects decrease with
increasing bond number in the usual way. The stationary state value is
$\lim_{\tau\to\infty} \sigma_N(\tau)/N(\tau) = ((1+\gamma) N_{eq})^{-1/2} =
(\gamma N_t)^{-1/2}$. Therefore larger rebinding does not only
increase the equilibrium number of bonds, but also decreases the size
of the fluctuations around $N_{eq}$. This leads to a narrow
distribution for large cluster under strong rebinding, with a small
probability of coming close to the lower boundary.

\FIG{fig:VF_ReflOccu}%
{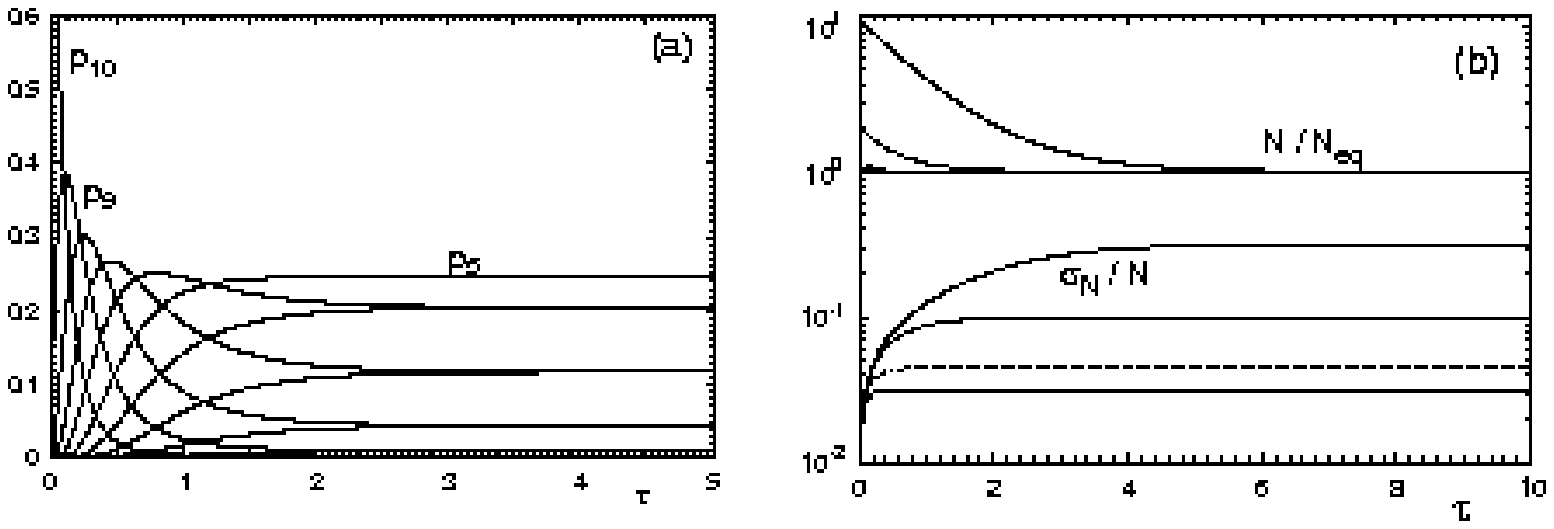}%
{(a) State probabilities $p_i(\tau)$ from \eq{eq:VF_states} for $N_t =
10$ with $f = 0$ and $\gamma = 1$ for a reflecting boundary at $i =
0$. (b) $N/N_{eq}$ and $\sigma_N/N$ for $N_t = 100$, $f = 0$ and
$\gamma = 0.1$, $1$, $5$ and $10$.}

In \fig{fig:VF_ReflOccu}a, we plot the state probabilities $p_i$ from
\eq{eq:VF_states} for cluster size $N_t = 10$ and rebinding
constant $\gamma = 1$. The system quickly relaxes to the equilibrium
state.  The only difference for different initial conditions is in the
initial transient. In particular, for $N_0 = N_{eq}$, the average does
not change in time, although the distribution initially spreads to the
binomial one. For $\gamma = 1$, the stationary distribution is
symmetric around the average. The width of the distribution for
different $\gamma$ is illustrated in \fig{fig:VF_ReflOccu}b, which
shows the average number of closed bonds normalised by the equilibrium
number of bonds, that is $N/N_{eq}$, together with the relative 
standard deviation, $\sigma_N/N$, for different values of the
rebinding constant $\gamma$. The curves for $N$ are independent of
$N_t$ due to the normalization. \fig{fig:VF_ReflOccu}b shows that with
increasing $\gamma$, relaxation becomes faster and the width of the
distribution decreases.

\FIGTIGHT{fig:VF_Norbits}%
{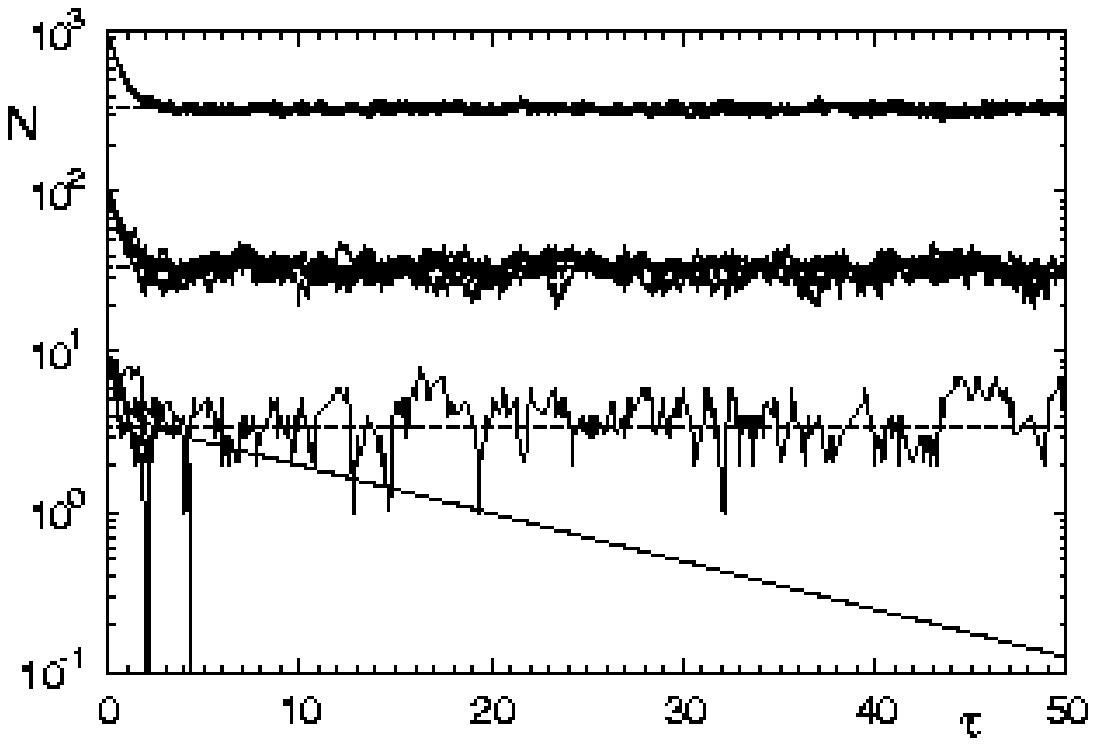}%
{Single simulation trajectories for $f = 0$, $\gamma = 0.5$, $N_t =
10, 100$ and $1000$ and an absorbing boundary at $i = 0$. Solid lines
are the average number of closed bonds $N$ and dashed lines are the
equilibrium number of closed bonds $N_{eq}$.}

For the biologically important case of an absorbing boundary at $i =
0$, it seems to be rather difficult to find a closed-form analytical
solution for arbitrary cluster sizes. For the case $N_t = 2$, we will
present such a solution in the next section.  For arbitrary $N_t$, we
use Monte Carlo simulations as described in
\sec{sec:MasterEquation}. In \fig{fig:VF_Norbits}, we show individual
simulation trajectories for different parameter values of interest, in
comparision to the average number of closed bonds for reflecting and
absorbing boundaries at $i = 0$. The plots show that the number of
closed bonds in a cluster first relaxes towards the steady state
value, for which rupture and rebinding balance each other.  Although
for the absorbing boundary the number of closed bonds decreases with
time in average, for individual realizations it stays roughly
constant, until a large fluctuation towards the absorbing boundary
leads to loss of this realization. The time-scale for the decrease in
$N$ is thus determined by the probability for fluctuations from the
steady state to the absorbing boundary. 

\FIG{fig:VF_occupancy}%
{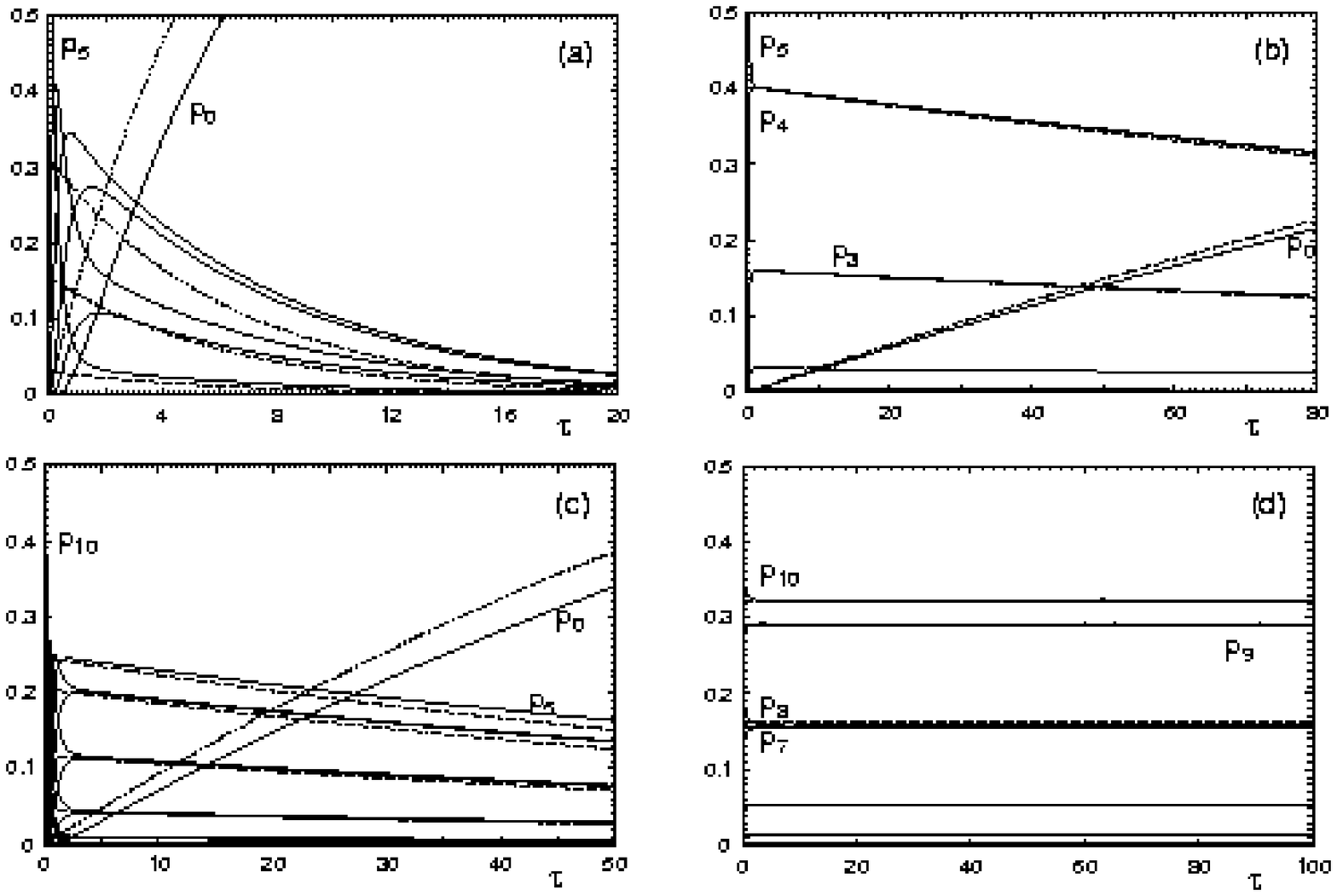}
{State probabilities $p_i$ as a function of time $\tau$ for different
cluster sizes $N_t = 5$ (a and b) and $N_t = 10$ (c and d) and for
rebinding rates $\gamma = 1.0$ (a and c) and $5.0$ (b and d). The
numerical solutions (solid curves) are compared to the leakage
approximation \eq{eq:LTEc} (dashed curves).}

Because a full analytical solution is not available for the case of an
absorbing boundary, we now introduce an approximation for this
case. It is similar to the local thermal equilibrium description
introduced by Zwanzig for modelling protein folding dynamics
\cite{c:zwan95}. Our starting point is that for large clusters and
strong rebinding, the absorbing boundary is a small perturbation to
the solution for the reflecting boundary, \eq{eq:VF_states}, which in
the following we will denote by $\{\bar p_i\}_{i=0}^{N_t}$. Since
$g(0) = 0$ for the absorbing boundary, $\dot p_0 = r_1 p_1$ with $r_1
= 1$ and probability will only accumulate in the completely
dissociated state. Since $p_0$ is slaved to the other state
probabilities and since we expect only a small perturbation for the
states with $i \geq 1$, we assume that here the different boundary
only leads to a simple renormalization caused by the 'leakage' into
the absorbing boundary:
\begin{align} \label{eq:LTEa}
p_i(\tau) & = \bar p_i(\tau) \left(1 - p_0(\tau)\right) \quad\text{for}\quad i \geq 1 \\
p_0(\tau) & = \int_0^{\tau} p_1(\tau') d\tau'\ . \nonumber 
\end{align}
Since relaxation to the steady state is faster than
decay to the absorbing boundary, $\bar p_i(\tau)$ can be taken to be
the stationary value, that is the constant $\bar p_i(\infty)$
according to \eq{eq:binomial}. Then $p_0(\tau) = \bar p_1(\infty)
\int_0^{\tau} (1-p_0(\tau')) d\tau'$, which is solved by $p_0(\tau) =
1 - e^{-\bar p_1(\infty) \tau}$.  Therefore \eq{eq:LTEa} simplifies to
\begin{align} \label{eq:LTEc}
p_i(\tau) &= \bar p_i(\infty) e^{-\bar p_1(\infty)\tau} \quad\text{for}\quad i \geq 1 \\
p_0(\tau) &= 1 - e^{-\bar p_1(\infty) \tau}\ . \nonumber
\end{align}
We conclude that the solution decays exponentially on the time scale
$1 / \bar p_1(\infty)$. In \fig{fig:VF_occupancy}, we plot Monte Carlo
solutions for the state probabilities in comparison to the
approximation. For $\gamma = 1$, the approximation does not work well
for $N_t = 5$, but it does so already for $N_t = 10$.  For $\gamma =
5$, the approximation works well for both cluster sizes. Note that in
this approximation, a term $\bar p_0(\infty) e^{-\bar
p_1(\infty)\tau}$ is missing for proper normalization
$\sum_{i=0}^{N_t} p_i = 1$. This is a small error for large clusters
and strong rebinding. In order to assess the validity of \eq{eq:LTEc},
we note that it presupposes that the time scale for relaxation to the
steady state, $1/(1+\gamma)$, is smaller than the time scale for decay
to the absorbing boundary, $1 / \bar p_1(\infty) =
(1+\gamma)^{N_t}/{\gamma N_t}$. Therefore $\gamma$ should be larger
than $(N_{eq})^{1/N_t}-1$.

\FIG{fig:VF_Naverage}%
{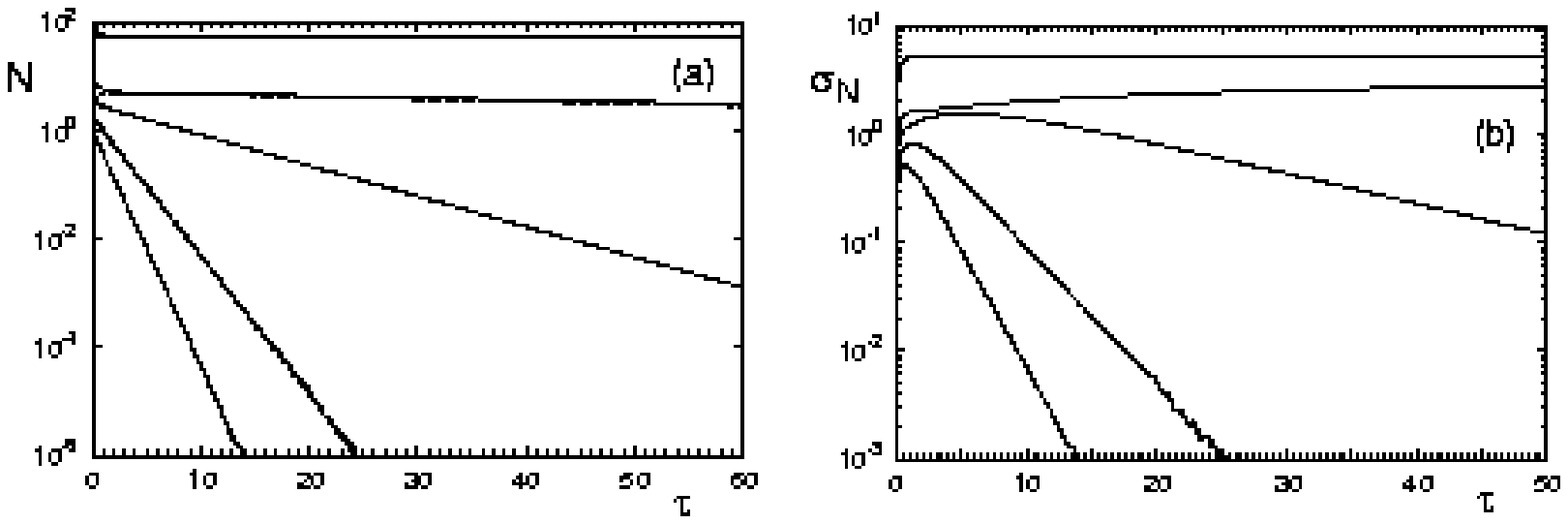}%
{(a) Average number of closed bonds $N$ obtained from stochastic simulations of 
the master equation for $\gamma = 1$ and $N_t = 1, 2, 5, 10$ and $100$. 
(b) Variance $\sigma_N$ of the cluster size distribution for the same
parameters as in (a).}

It follows from \eq{eq:LTEc} that the mean number of closed bonds
decay in an exponential way, $N(\tau) = N_{eq} e^{-\bar p_1(\infty)
\tau}$. This is confirmed by \fig{fig:VF_Naverage}a, which shows the
corresponding simulation results. For $N_t = 2, 5$ and $10$, we have
$\bar p_1(\infty) = 0.5, 0.16$ and $9.7\times10^{-3}$. Numerically we
find $0.6, 0.13$ and $0.01$, thus the approximation is rather good.
In \fig{fig:VF_Naverage}b, we plot numerical results for the standard
deviation $\sigma_N$. The initial increase of $\sigma_N$ is well
described by \eq{eq:VF_Nvariance} for the reflecting boundary, thus
the boundary has little influence here.  Large clusters stay close to
the steady state during the time shown and the approximation is
applicable. For small clusters, the variance grows larger than the
steady state value before is decreases exponentially while the cluster
size $N$ approaches zero. The variance contains two time-scales. The
second moment of the distribution decreases on the same timescale as
the average, while the square of the first moment decreases twice as
fast. The long time exponential decrease of $\sigma_N$ is thus
described by twice the relaxation time as is was found for the average
number of bonds.

Although an exact solution for the state probabilities seems to be
impossible for the case of an absorbing boundary, more analytical
progress can be made if one is only interested in the probability that
the cluster dissociates as a whole. For the absorbing boundary, the
cluster dissociation rate has been denoted by $D(\tau)$ before. For
the reflecting boundary, $D(\tau)$ can be identified as the
probability that the state $i = 0$ is reached for the first time at
time $\tau$ if the system has started in the state $i = N_t$ at time
$\tau = 0$. This is a first passage problem which can be treated with
Laplace techniques. Since the transition rates do not depend on
absolute time, one can decompose the state probability for $i = 0$
into two parts:
\begin{equation}        
p_0(\tau) = \int_0^\tau D(\tau') p_{0,0}(\tau - \tau') d\tau'
\end{equation}
where $p_{0,0}(\tau)$ is the state probability for state $i = 0$ with
initial condition $p_i(0) = \delta_{i,0}$.  $p_{0,0}(\tau)$ can also
be interpreted as the probability for having returned to the boundary
after time $\tau$. A Laplace transform of the equation leads to an
algebraic relation between the Laplace transforms of the three
functions:
\begin{equation}
D(s) = \frac{p_0 (s)}{p_{0,0}(s)}\ .
\end{equation}
Here $D(s) = \int_0^{\infty} e^{-s\tau} D(\tau)$ denotes the
Laplace transform of the function $D(\tau)$. The explicit form of the 
probability $p_0(\tau)$ is given in \eq{eq:VF_states}: 
\begin{equation}
p_0(\tau) = \left( \frac{1 - e^{-(1+\gamma)\tau}}{1+\gamma}\right)^{N_t}\ .
\end{equation}
The probability $p_{0,0}(\tau)$ can also be calculated with standard
techniques \cite{b:goel74}:
\begin{equation}
p_{0,0}(\tau) = \left( \frac{1 + \gamma e^{-(1+\gamma)\tau}}{1+\gamma}\right)^{N_t}\ .
\end{equation}
The Laplace transforms of the these two functions are given by 
\begin{align}
p_0(s) &= \frac{1}{\left(1+\gamma\right)^{N_t}}
       \sum_{i = 0}^{N_t}\binom{N_t}{i}\frac{(-1)^i}{s + i (1+\gamma)}\\
\intertext{and}
p_{0,0}(s) &= \frac{1}{\left(1+\gamma\right)^{N_t}}
           \sum_{i = 0}^{N_t}\binom{N_t}{i}\frac{\gamma^i}{s+i(1+\gamma)}\quad,\\
\intertext{so that the Laplace transformed first passage probability time distribution is}
\label{eq:fpt_laplace}
D(s) &= \frac{\sum_{i = 0}^{N_t}\binom{N_t}{i}\frac{(-1)^i}{s + i(1+\gamma)}} 
                            {\sum_{i = 0}^{N_t}\binom{N_t}{i}\frac{\gamma^i}{s+i(1+\gamma)}}\quad.
\end{align}
Unfortunately, the analytical backtransform for $D(s)$ seems to be impossible.
However, the mean first passage time can be extracted from this result,
because it does not require the backtransfrom \cite{b:hone90}:
\begin{equation} \label{eq:VF_lifetime}
T = \left.\frac{dD(s)}{ds}\right|_{s = 0}
  = \frac{1}{1+\gamma}\left(\sum_{n = 1}^{N_t}\left\{\binom{N_t}{n}\frac{\gamma^n}{n}\right\} + H_{N_t}\right)\ .
\end{equation}
This equation is a polynomial of order $N_t-1$ in $\gamma$. The zero
order term is the harmonic number $H_{N_t}$, so for $\gamma = 0$ we
recover the result from \eq{eq:harmonic_number}.  In
\fig{fig:VF_lifetime}, we plot \eq{eq:VF_lifetime} as a function of 
cluster size $N_t$ and rebinding rate $\gamma$. As long as $\gamma <
1$, cluster lifetime grows only weakly (logarithmically) with cluster
size (at least for not too large clusters).  For $\gamma > 1$, the
higher order terms in $\gamma$ take over and the increase in $T$
becomes effectively exponential, as shown in
\fig{fig:VF_lifetime}a. In \fig{fig:VF_lifetime}b, it is shown
explicitly that increasing $\gamma$ to values larger than unity leads
to a strong increase in lifetime.  This effect is larger for larger
clusters since the number of rebinding events in the dissociation path
is larger.

\FIG{fig:VF_lifetime}%
{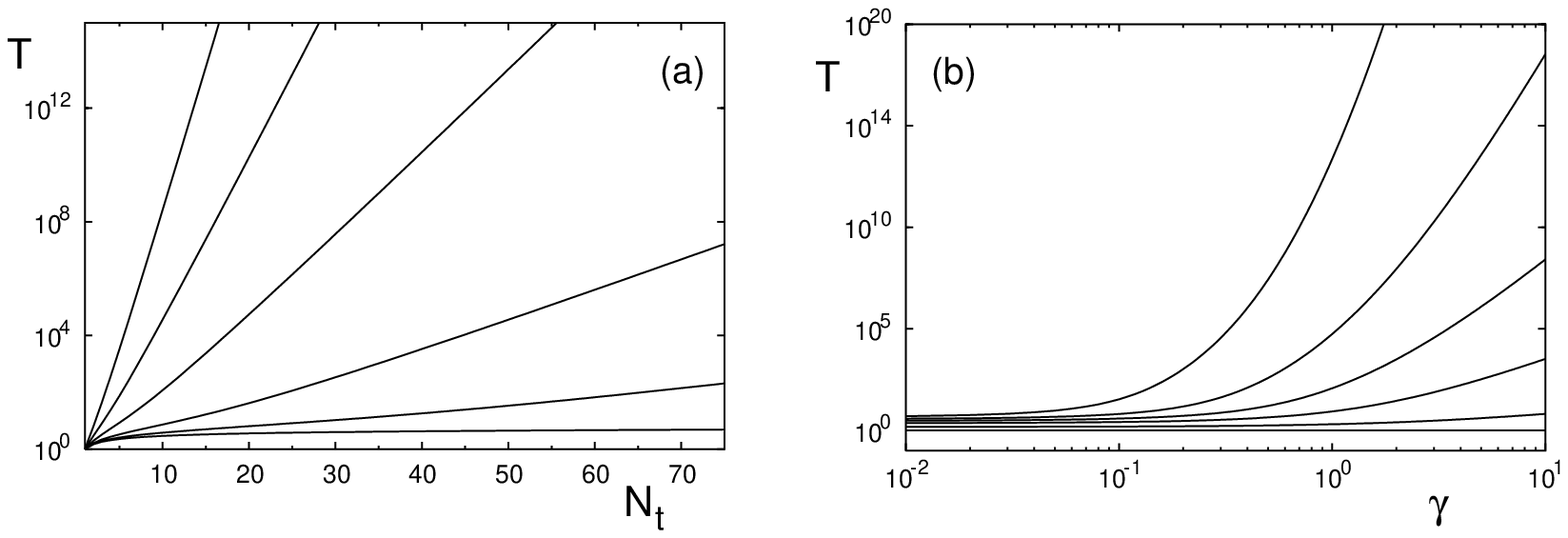}%
{(a) Average lifetime $T$ of adhesion clusters as function of cluster
size $N_t$ for rebinding rates $\gamma = 0.0, 0.1, 0.5, 1, 5$ and
$10.0$. (b) Average lifetime $T$ as function of rebinding constant
$\gamma$ for cluster size $N_t = 1, 2, 5, 10, 20$ and $50$.}

\section{Finite force and finite rebinding}
\label{sec:FR}

\subsection{Deterministic analysis} 

Force destabilizes the cluster, while rebinding stabilizes it again.
It has been shown by Bell that in the framework of the deterministic
equation \eq{DeterministicEquation}, the cluster remains stable up to
a critical force $f_c$ \cite{c:bell78}. For the following it is
helpful to revisit his stability analysis. In equilibrium we have
\begin{equation} \label{eq:equilibrium}
N_{eq} e^{f / N_{eq}} = \gamma (N_t - N_{eq})\ .
\end{equation}
At small force $f$, this equation has two roots, with the larger one
corresponding to a stable equilibrium. As force increases, a
saddle-node bifurcation occurs. Above the critical force, no roots
exist and the cluster becomes unstable. Exactly at critical loading,
the two roots collapse and the slopes of the two terms become equal.
This gives an additional equation
\begin{equation} 
e^{f_c / N_c} (1 - \frac{f_c}{N_c}) = - \gamma\ .
\end{equation}
These two equations allow to determine the critical values for cluster
size and force:
\begin{equation}\label{eq:fcrit}
f_c      = N_t\ \rm plog\left(\frac{\gamma}{e}\right) \quad\text{and}\quad
N_{c} = N_t\ \frac{\rm plog\left(\frac{\gamma}{e}\right)}%
                     {1 + \rm plog \left(\frac{\gamma}{e}\right)}\quad,
\end{equation}
where the product logarithm $\rm plog(a)$ is defined as the solution
$x$ of $xe^x=a$. For small forces, the unstable fixed point is very
close to zero. This implies that the stable fixed point is an
attractor for most initial conditions. Close to the critical force,
the unstable fixed point is close to the stable one and only the
initial conditions above $N_c$ will reach the stable fixed
point. \eq{eq:fcrit} scales in a trivial way with $N_t$, but in a
complicated way with $\gamma$. For $\gamma < 1$, we have $f_c
\approx \gamma N_t / e$. Thus the critical force vanishes
with $\gamma$, because the cluster decays by itself with no
rebinding. For $\gamma > 1$ and up to $\gamma \approx 100$, we have
$f_c \approx 0.5 N_t \ln \gamma$.  This weak dependence on $\gamma$
shows that the single bond force scale set by $F_b$ also determines
the force scale on which the cluster as a whole
disintegrates. \fig{fig:fcrit}a shows how $f_c$ crosses over from
linear to logarithmic scaling with $\gamma$.

\FIG{fig:fcrit}%
{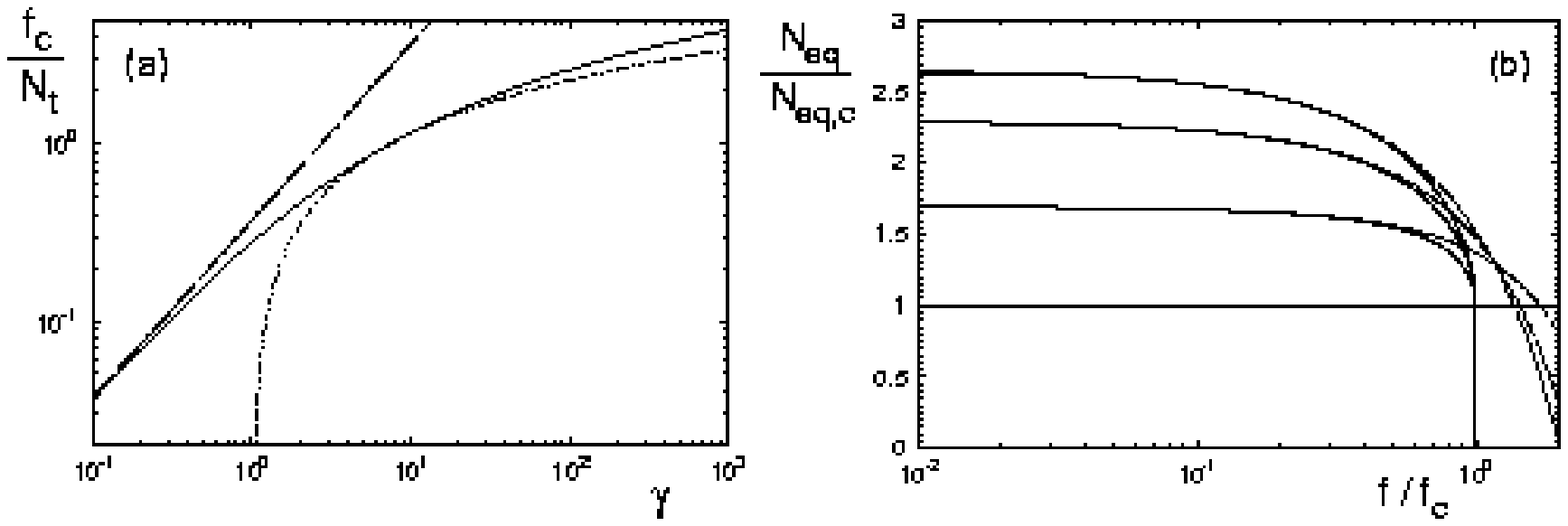}%
{(a) Critical force \eq{eq:fcrit} in relation to total cluster size,
$f_c/N_t$, as function of the rebinding constant $\gamma$. It scales
linearly (dotted curve) at small and logarithmically (dashed curve) for
larger $\gamma$. (b) Stable steady value for number of closed bonds
normalised by the critical value, $N_{eq}/N_{eq,c}$, for $\gamma =
0.1, 1$ and $10$ as function of force $f/f_c$ for $N_t =
100$. Numerical results (solid lines) are compared to the
approximation \eq{eq:Nfix_small} (dashed lines). The horizontal line
marks the smallest possible value at the critical force $f_c$.}%

\eq{eq:equilibrium} is an implicit equation which cannot be inverted to 
give $N_{eq}$ as a function of the model parameters $N_t$, $f$ and
$\gamma$. In general, $N_{eq}$ decreases from $\gamma N_t/(1+\gamma)$
for $f = 0$ to $N_{c}$ for $f_c$. For small forces we can find an
approximate solution by first expanding the exponential function in
\eq{eq:equilibrium} to second order in $f/N_{eq}$ and then expanding
the resulting quadratic function for $N_{eq}$ to second order in
$f/\gamma N_t$:
\begin{equation}\label{eq:Nfix_small}
N_{eq} \approx \frac{\gamma N_t}{1+\gamma} \left[ 1 - \left( \frac{f}{\gamma N_t} \right) 
               - \frac{\gamma + 1}{2} \left( \frac{f}{\gamma N_t} \right)^2 \right]\quad .
\end{equation}
\fig{fig:fcrit}b shows numerical results for $N_{eq}/N_{c}$ in
comparison with the low force approximation \eq{eq:Nfix_small} for
different rebinding constants $\gamma = 0.1, 1.0$ and $10$ as function
of force $f/f_c$ and for cluster size $N_t = 100$. 

\FIGTIGHT{fig:lifetime}%
{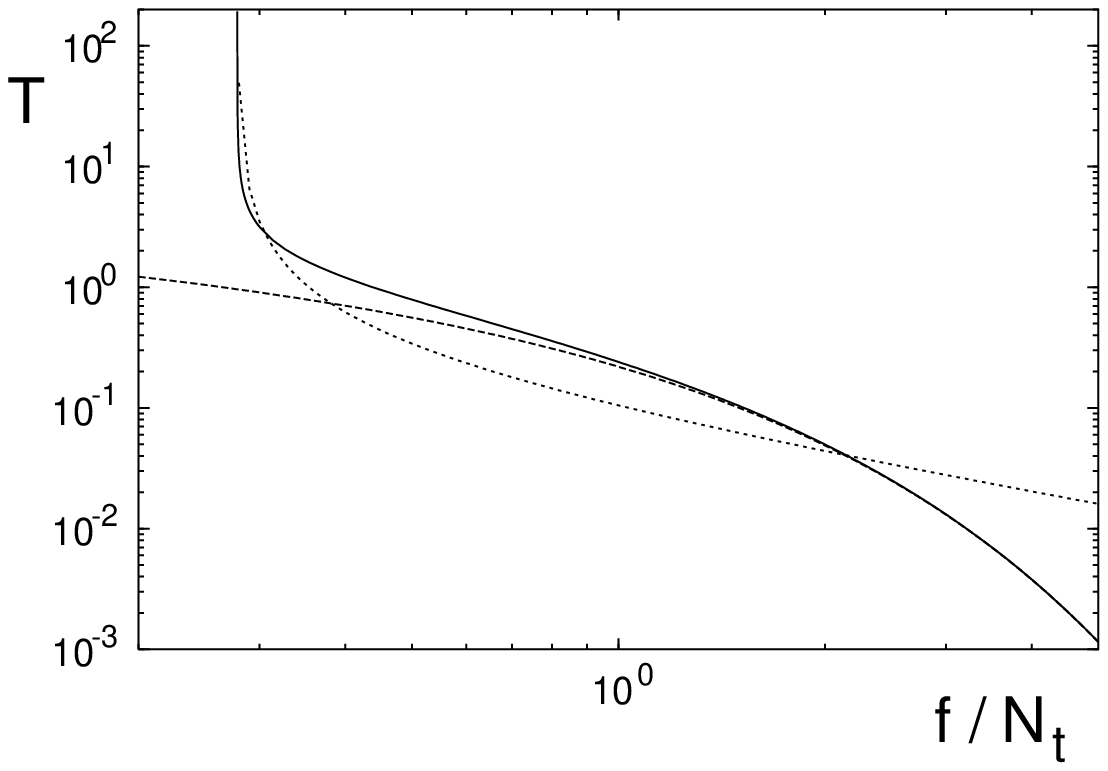}%
{Deterministic mean lifetime $T$ of a cluster of $N_t = 100$ bonds for
$\gamma = 1$ as a function of the force-size ratio $f/N_t$. Numerical
integration of the deterministic equation (solid line) is compared to
the exact solution for vanishing rebinding (dashed line) and the
inverse linear scaling (dotted line) predicted close to the critical
force.}%

In the deterministic framework, cluster lifetime is infinite for $f <
f_c$, because a stationary state exists at $N_{eq}$. For $f > f_c$,
cluster lifetime is finite, but strongly varies as a function of
$N_t$, $f$ and $\gamma$. For $f \gg f_c$, we can neglect rebinding and
use the results from \sec{sec:VR}, where we found for cluster lifetime
\begin{equation} \label{eq:Tdetlarge}
T_{det} = E\left( \frac{f}{N_t} \right) \approx \frac{e^{- (f / N_t)}}{1 + (f / N_t)}\ ,
\end{equation}
compare \eq{eq:VR_Tdetlarge}. As force $f$ is decreased from above
towards the critical value $f_c$, rebinding becomes important again
and cluster lifetime diverges. To understand this limit, we note that
here the system will evolve very slowly, because it is still close to
a steady state. Therefore we can expand the time derivative of $N$,
compare \eq{DeterministicEquation}, for small deviations from the
critical state:
\begin{equation} \label{eq:SlopeLinear}
\frac{dN}{d\tau} \approx \left. \frac{\partial}{\partial f}\left(\frac{dN}{d\tau}
\right)\right|_{f_c,N_c}\left(f-f_{c}\right)  
= - e^{f_c / N_c} (f - f_c) = - e^{1 + \rm plog(\gamma / e)} (f - f_c)\ .
\end{equation}
In this limit, the lifetime will be dominated by the time spent close
to the critical state. The time for a significant change $\Delta N
\simeq -1$ in $N$ is
\begin{equation}\label{eq:Tcrit}
T \approx \Delta \tau \approx  e^{- (\rm plog(\gamma / e) + 1)} \frac{1}{f - f_c}\ .
\end{equation}  
Therefore $T$ diverges like the inverse of $f -
f_c$. \fig{fig:lifetime} shows the lifetime of an adhesion cluster
derived from numerical integration of the deterministic equation for
$\gamma = 1$ and $N_t = 100$ as a function of the force-size ratio $f
/ N_t$. The numerical results are compared to the approximation
\eq{eq:Tdetlarge} for large forces and \eq{eq:Tcrit} for the divergence 
close to the critical point.  Obviously both approximations work well
for their respective limits.  For different cluster sizes $N_t$, the
plot remains basically unchanged (not shown), because the forces above
$f_c$ are already in the range where \eq{eq:VR_Tdet} predicts scaling
with $f/N_t$ alone. For different rebinding constants $\gamma$ the
results are qualitatively the same, only that the critical force is
shifted to different values.

\subsection{Stochastic analysis}

In general, it seems to be difficult to find a closed-form analytical
solution for the state probabilities $p_i(\tau)$ for general values of
$\gamma$, $f$ and $N_t$. In the following, we will derive such an
analytical solution for the case $N_t = 2$ with an absorbing boundary.
In principle, solutions can be constructed in the same way for
a reflecting boundary or larger clusters, but for increasing cluster
size, the analytical procedure quickly becomes intractable. For this
reason, we will later use simulations to deal with the general case.

We start by rewriting the master equation \eq{MasterEquation} in matrix form:
\begin{equation} \label{eq:W_meq}
\dot {p} = {W} \cdot {p}\ .
\end{equation}
For the case $N_t = 2$ with an absorbing boundary, ${p} =
(p_0,p_1,p_2)^T$ and
\begin{equation}\label{eq:W}
{W} = \begin{pmatrix} 0 &   r_1        &    0 \\ 
                             0 & -(r_1 + g_1) &  r_2 \\ 
                             0 &         g_1  & -r_2 \end{pmatrix}\ .
\end{equation}
\eq{eq:W_meq} is solved by \cite{b:hone90}
\begin{equation}
{p}(\tau) = e^{{W} \tau} \cdot {p}(0)
= \sum_{\lambda} c_{\lambda} e^{\lambda\tau} {p}_{\lambda}
\end{equation}
where $\lambda$ and ${p}_{\lambda}$ are the eigenvalues and
eigenvectors of ${W}$, respectively. The coefficients
$c_{\lambda}$ have to be determined from the initial condition
\begin{equation}\label{eq:initial}
{p}(0) = \sum_{\lambda} c_{\lambda} {p}_{\lambda}\ .
\end{equation}
Since the absorbing state ${p}_{0} = (1,0,0)^T$ is a stationary
state, the corresponding eigenvalue $\lambda_0 = 0$. The other two eigenvalues
are negative and correspond to transient states:
\begin{equation}\label{eq:EV}
\lambda_{1,2} = - (\Omega \pm \omega) 
\end{equation}
with $\Omega$ and $\omega$ being defined as 
\begin{equation} \label{eq:definitionOmega}
\Omega = (r_1 + r_2 + g_1) / 2 \quad\text{and}\quad \omega = \sqrt{\Omega^2 - r_1 r_2}\ .
\end{equation}
Note that $0 < \omega < \Omega$ and hence $\lambda_{1,2} < 0$. The transient eigenstates are 
\begin{equation}\label{eq:ES}
{p}_{1} = \frac{1}{g_1}\begin{pmatrix} \lambda_2 + r_1 \\ \lambda_1 + r_2 \\ g_1\end{pmatrix}
\quad\text{and}\quad
{p}_{2} = \frac{1}{g_1}\begin{pmatrix} \lambda_1 + r_1 \\ \lambda_2 + r_2 \\ g_1\end{pmatrix}\ .
\end{equation}
The three eigenstates ${p}_{\lambda}$ are linearly
independent and form a basis of the state space of the cluster.  
With the initial condition $p_i = \delta_{i,N_t}$, that is
${p}(0) = (0,0,1)^T$, the coefficients $c_{\lambda}$ follow
from \eq{eq:initial} as 
\begin{equation}
c_{\lambda_0} = 1 \ , \quad c_{\lambda_1} = \frac{\lambda_2 +
r_2}{\lambda_2 - \lambda_1} \quad\text{and}\quad c_{\lambda_2} =
-\frac{\lambda_1 + r_2}{\lambda_2 - \lambda_1} \ .
\end{equation}
The final result then can be written as
\begin{eqnarray}
p_0(\tau) &=& 1 - \left[ \cosh(\omega\tau) 
+ \frac{\Omega}{\omega} \sinh(\omega\tau) \right] e^{-\Omega\tau}\ , \nonumber \\
\label{eq:two_bonds_exact}
p_1(\tau) &=&     \frac{r_2}{\omega} \sinh(\omega\tau) e^{-\Omega\tau}\ ,  \\
p_2(\tau) &=&     \left[ \cosh(\omega\tau) 
+ \frac{\Omega-r_2}{\omega}\sinh(\omega\tau)\right] e^{-\Omega\tau}\ . \nonumber
\end{eqnarray}
There is a competition between the hyperbolic terms, which grow on
the time-scale $1/\omega$, and the exponential terms, which decrease on
the time-scale $1/\Omega$. Since $\omega < \Omega$, the exponential
terms will win and only the stationary state survives.

With the exact solution \eq{eq:two_bonds_exact}, one now can calculate
any quantity of interest. For example, the mean number of bonds,
$N(\tau) = \sum_i i p_i$, follows as
\begin{equation}
N(\tau) = \left[2 \cosh(\omega\tau) + \frac{r_1 
+ g_1}{\omega} \sinh(\omega\tau)\right] e^{-\Omega\tau} \ .
\end{equation}
The dissocation rate for the cluster as a whole as given by
$D(\tau) = r_1 p_1 = \dot p_0$, resulting in 
\begin{equation}
D(\tau) =  \frac{r_1 r_2}{\omega}\sinh(\omega\tau)e^{-\Omega\tau}\ . 
\end{equation}
One easily checks that normalization is correct, $\int_0^{\infty}
D(\tau) d\tau = 1$.  Mean cluster lifetime $T$ now follows as
\begin{equation} \label{eq:T_two_bonds_exact}
T = \int_0^{\infty} \tau D(\tau) d\tau 
= \frac{1}{2} \left( 2 e^{-f} + e^{-f/2} + \gamma e^{-3f/2} \right)\ .
\end{equation}
As shown in the preceding sections for special cases,
force leads to exponentially decreased lifetimes, while rebinding
leads to polynomial terms in $\gamma$ up to order $N_t - 1$.

\FIG{fig:FR_occupancy}%
{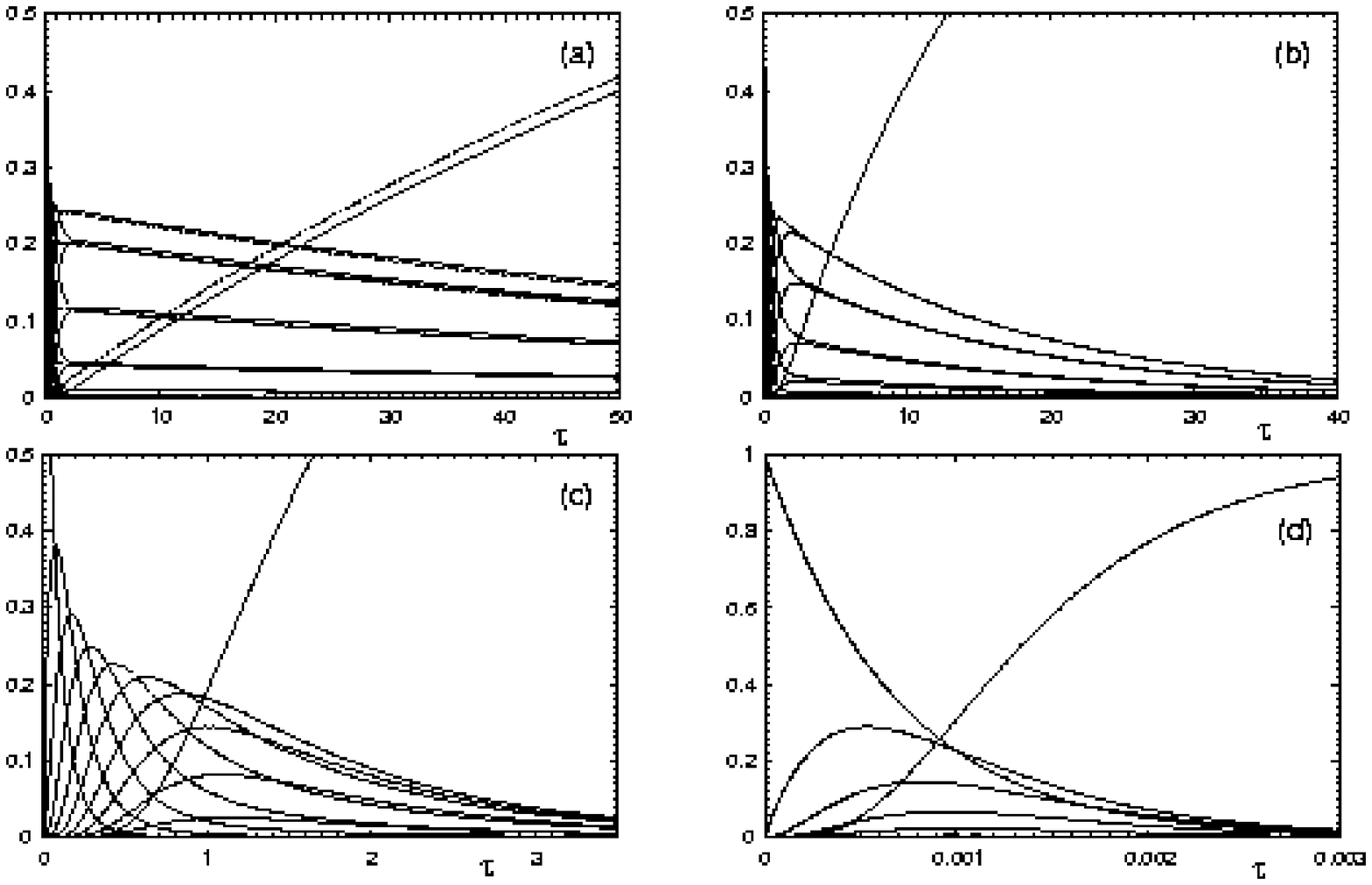}%
{State probabilities $p_i$ as a function of time $\tau$ for $N_t = 10$
with $\gamma = 1$ for force $f = 0.1, 1, 3$ and $10$ (a - d).  In (a),
the numerical solution is compared with the leakage approximation
(dotted lines). The intermediate force values are chosen below and
above the critical force $f_c = 0.278 N_t$.}%

Although the eigenvalue analysis can be used also for the general case
of arbitrary cluster size, in this case it is more efficient to use
exact stochastic simulations as described in
\sec{sec:Master_equation_numerical}.  In \fig{fig:FR_occupancy}
numerical solutions of the state occupancy probabilities
$\{p_i\}_{i=0}^{N_t}$ are plotted for $N_t = 10$ with $\gamma = 1$ for
four different forces $f = 0.1, 1, 3$ and $50$. This figure
corresponds to \fig{fig:VR_occupancy} for vanishing rebinding and
\fig{fig:VF_occupancy} for vanishing force. For small force, the
numerical solutions compare well with the approximation \eq{eq:LTEc}
introduced for vanishing force. For larger force, but still below the
critical force, the state probabilities still decrease exponentially
for large times, but the approximation \eq{eq:LTEc} breaks down,
because the reference distribution $\{\bar p_i(\infty)\}_{i =
0}^{N_t}$ now had to be replaced by the unknown steady state for the
case of finite force. If force is increased beyond the critical force
($f_c = 2.78$ for $\gamma = 1$), a simple description is not
available, because equilibration and decay occur on the same
timescale.  For very large force, the behavior of the adhesion
cluster approaches that for vanishing rebinding, with the analytical
solution \eq{eq:tees20}.

\FIG{fig:FR_Norbits}%
{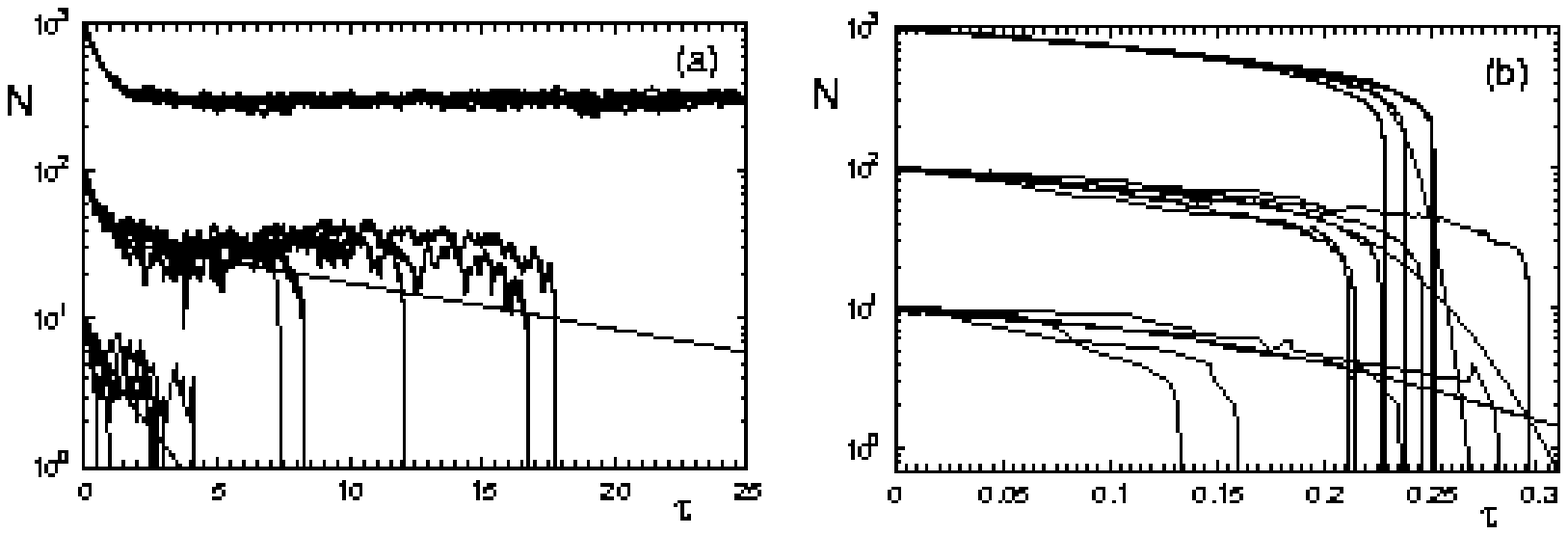}%
{Single simulation trajectories for $N_t = 10, 100$ and $100$ for
$\gamma = 1$ and at two different forces (a) $f = 0.25 N_t < f_c$ and
(b) $f = N_t > f_c$. Representative trajectories are compared to the
average number $N$ of closed bonds resulting from averaging over a large
number of such trajectories.}

\fig{fig:FR_Norbits} demonstrates that the decay process changes
dramatically as force is increased above the critical value.  It
displays trajectories of individual clusters with initially $N_t = N_0
= 10, 100$ and $1000$ closed bonds in comparison with the average
number of bonds derived from a large number of these
trajectories. Since $f_c = 0.278 N_t$ for $\gamma = 1$,
\fig{fig:FR_Norbits}a with $f = 0.25 N_t$ is below the critical value. 
For the largest cluster, the system equilibrates towards the steady
state and then fluctuate around this value with very rare encounters
of the absorbing boundary. For the smaller clusters, however,
fluctuations towards the absorbing boundary frequently lead to loss of
individiual realizations. As a result, the average number of closed
bonds decays exponentially on a much faster
timescale. \fig{fig:FR_Norbits}a with $f = N_t$ is above the critical
force and the behavior is changed qualitatively. A steady state does
not exist anymore and the clusters do not decay by fluctuations, but
the size of each adhesion cluster is continuously reduced. Clusters of
different size now decay on the same timescale and rebinding events
are very rare in comparison to rupture events.

\FIG{fig:FR_Naverage}
{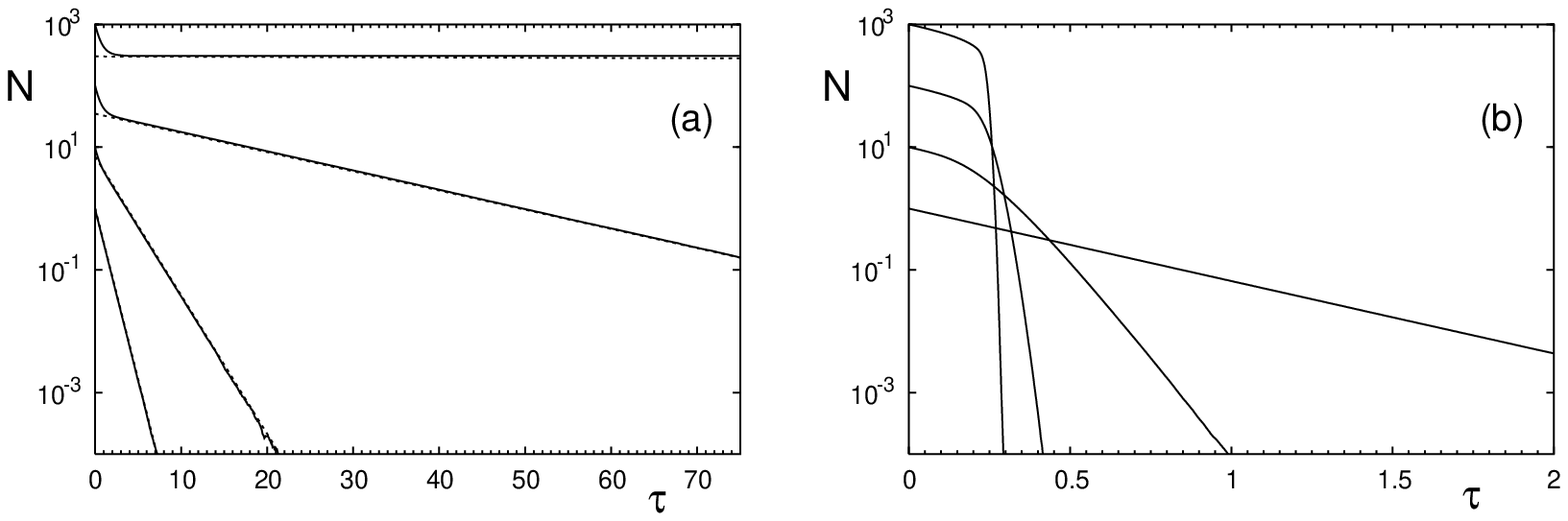}%
{Average number of closed bonds $N$ for $N_t = 1, 10, 100$ and $1000$,
$\gamma = 1$ and (a) $f/N_t = 0.25$ and (b) $f/N_t = 1$. In (a), the
numerical results are compared to exponentially decaying curves $\sim
e^{-a\tau}$ (dashed lines) with $a = 1.28, 0.52, 0.072$ and $0.0009$
for increasing bond number.} %

\fig{fig:FR_Naverage} plots numerical results for the average number of closed 
bonds $N$ as function of time $\tau$ for two different values of $f /
N_t$ and for cluster sizes $N_t = 1, 10, 10^2$ and $10^3$.  For $f =
0.25 N_t < f_c$, after initial relaxation all curves decay
exponentially.  For $f = N_t > f_c$, the larger clusters show a steep
decrease in average cluster size at the end of the decay due to the
effects of shared loading.  For the small clusters, the average
cluster size decreases slowly since cooperative effects are small.

\FIG{fig:FR_Ncumulants}
{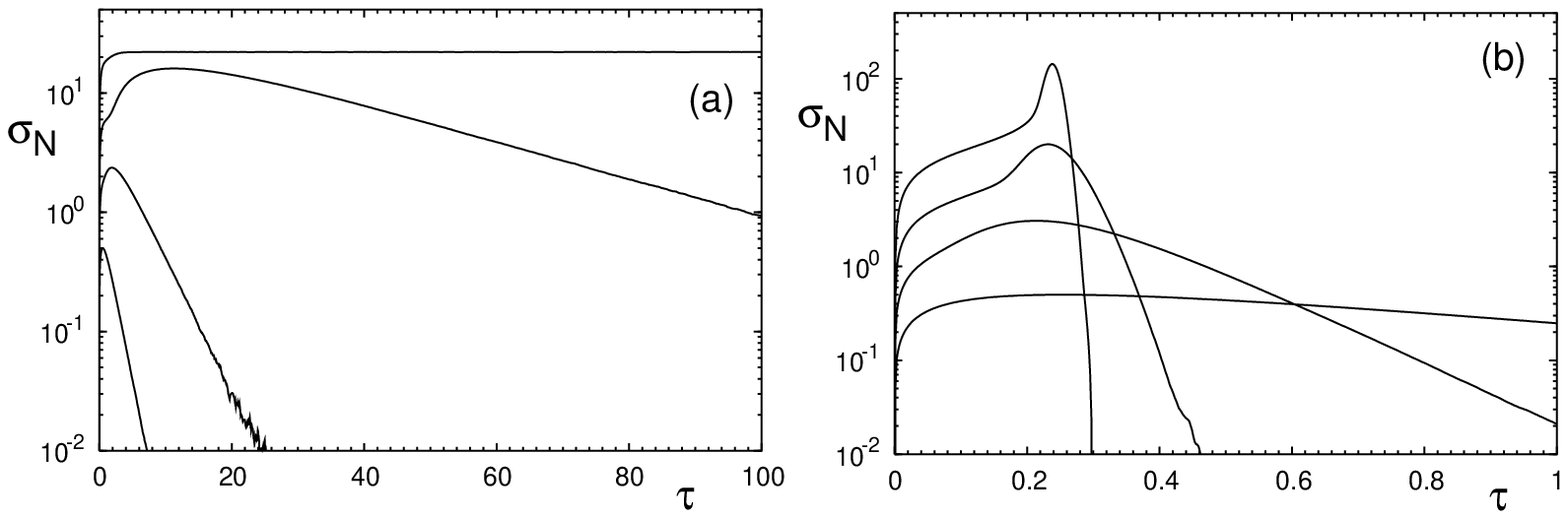}%
{Variance $\sigma_N$ for forces (a) $f = 0.25 N_t$ and (b) $f = N_t$
for $\gamma = 1$ and $N_t = 1, 10, 100$ and $1000$.}%

\fig{fig:FR_Ncumulants} plots the variance $\sigma_N(\tau)$ of the
distribution $p_i$ for the two force values used in the two previous
figures. Below the critical force the behavior is similar to that for
vanishing force depicted in \fig{fig:VF_Naverage}. The variance
decreases exponentially after having traversed a maximum. For forces
above the critical force, a different behavior arises. After growing
as expected in the initial phase, the variance displays a sharp peak.
This effects becomes more pronounced the larger cluster size.

\FIG{fig:FR_NTcomp}%
{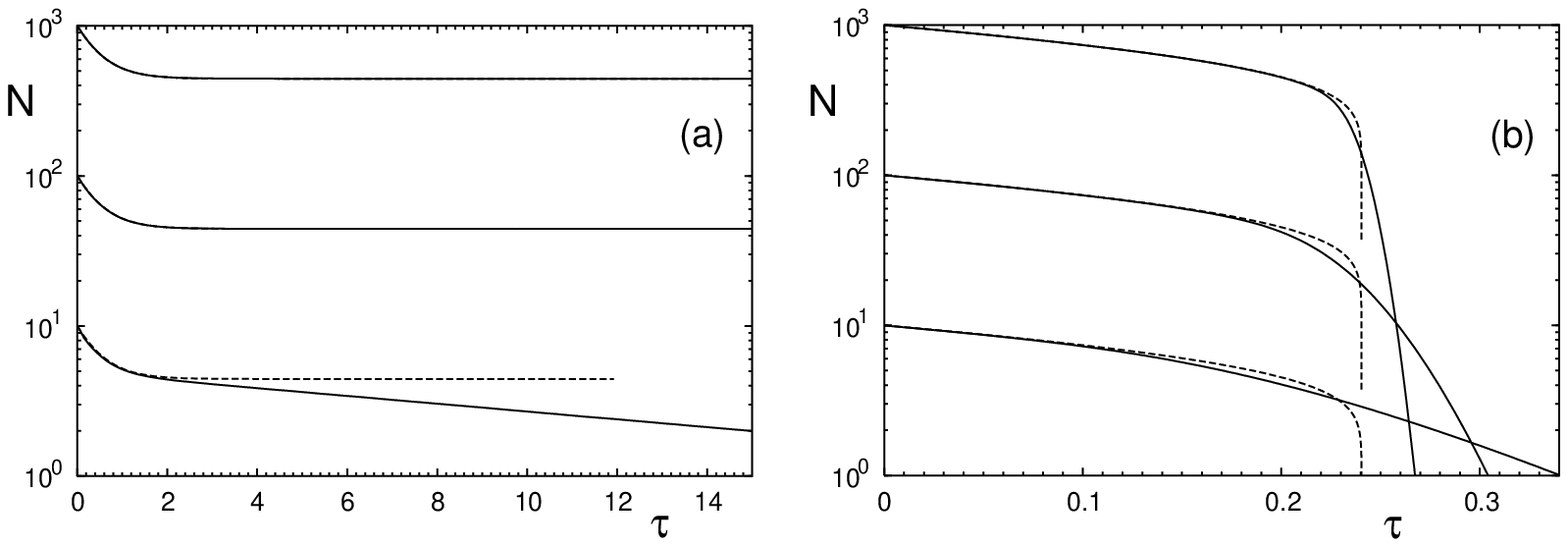}%
{Comparison of stochsatic and deterministic results for the average
number of closed bonds $N$ derived from numerical solutions of the
master equation (solid lines) and from integration of the
deterministic equation (dashed curves) for cluster sizes $N_t = 10,
100$ and $1000$ and forces (a) $f/N_t = 0.1$ and (b) $1.0$. The
rebinding rate is $\gamma = 1$.}

In \fig{fig:FR_NTcomp} a comparison of the average number of closed
bonds in the stochastic and the determinsitic description is
shown. $N(\tau)$ is plotted for cluster sizes $N_t = 10, 100$ and
$1000$ for the forces $f/N_t = 0.1$ and $1.0$ and the rebinding rate
$\gamma = 1.0$.  For small forces $f < f_c$, the average number of
closed bonds equilibrates towards the steady state and remains
constant thereafter. The fluctuations occuring in the stochastic
description lead to a slow decrease of $N$. Above the critical force,
the deterministic clusters decay as well, and in a more
abrupt way than the stochastic average.

\FIG{fig:FR_Tdissstribution}
{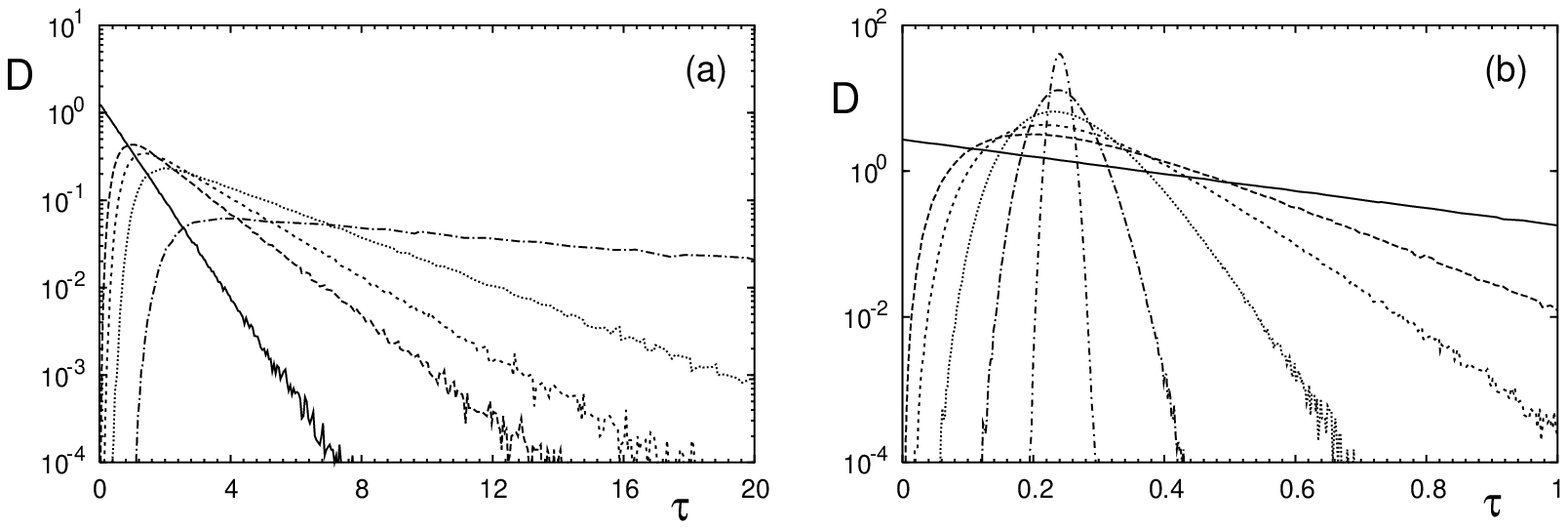}
{Dissociation rate $D$ of the overall cluster for $\gamma = 1$ and
$N_t = 1, 5, 10, 25, 100$ and $1000$. (a) $f = 0.25 N_t$ and (b) $f =
N_t$.}

We now turn to the dissociation rate of the overall cluster as a
function of the model parameters.  For $\gamma = 1$ and $f = 0.25 N_t$
and $N_t$, that is below and above the critical force, numerical
results are plotted in \fig{fig:FR_Tdissstribution}.  For a single
bond, dissociation is a Poisson process with the maximum at $\tau = 0$
and an exponentially decreasing dissociation rate $D = r(1) p_1 = e^f
e^{-e^f\tau}$.  For larger clusters and below the critical force,
fluctuations to the absorbing boundary determine the rate of
dissociation, which vanishes at $\tau = 0$, goes through a maximum and
then decreases exponentially with time. As explained above, the
exponential decay follows because decay proceeds by rare fluctuations
from the steady state towards the absorbing boundary.  Above the
critical force, the dissociation rate for $N_t > 1$ becomes more
sharply peaked and cannot be described with single exponential
curves. A steady state does not exist anymore and dissociation does
not proceed by fluctuations. The trajectories in \fig{fig:FR_Norbits}
have shown that adhesion clusters decay fairly abrupt towards the end
of the decay as a consequence of shared loading. This cooperative
instability is the reason for the sharp dissociation distribution for
large clusters under super-critical loading. The single bond that
lacks these cooperativity, still shows the exponential dissociation
rate which is now the slowest decaying for the given force size ratio.

Whereas results for the dissociation rate have to be obtained
numerically, the average lifetime can be calculated analytically
\cite{b:kamp92}.  The basic idea here is to sum the average times for
any possible pathway leading from the initial cluster size $N_0$
towards dissociation at the absorbing boundary $i = 0$ with its
appropriate statistical weight. One can show that the lifetime
$T_{N_t,N_0}$ of a cluster with a total of $N_t$ molecular bonds of
which $N_0$ are closed initially satisfies the equation \cite{b:kamp92}
\begin{equation}
g(N_0)\left(T_{N_t,N_0+1}-T_{N_t,N_0} \right) 
          + r(N_0)\left(T_{N_t,N_0-1}-T_{N_t,N_0}\right) = -1\ .
\end{equation}
The left hand side can be considered to be the adjoint operator of the
master equation acting on the average lifetime $T_{N_t,N_0}$.  For the
initial condition $N_0 = N_t$, the equation is solved exactly by
\begin{equation} \label{eq:FR_lifetime}
T = T_{N_t,N_t} = \sum_{i = 1}^{N_t} \frac{1}{r(i)} +
\sum_{i = 1}^{N_t - 1} \sum_{j = i+1}^{N_t} 
\frac{\prod_{k = j-i}^{j-1} g(k)}{\prod_{k = j-i}^{j} r(k)}\ ,
\end{equation}
where the first term is the result \eq{eq:lifetime} for vanishing
rebinding and the second term results in a polynomial of order $N_t -
1$ in $\gamma$. For $f = 0$, \eq{eq:FR_lifetime} is identical to the
earlier result \eq{eq:VF_lifetime} obtained by Laplace transforms.
Both expressions are polynomials of order $N_t - 1$ in $\gamma$, but
in the general case from \eq{eq:FR_lifetime}, the coefficients depend
on force. For $N_t = 2$, we obtain the result from
\eq{eq:T_two_bonds_exact}. For $N_t = 3$, we find
\begin{equation}
T = e^{-f} + \frac{e^{-f/2}}{2} + \frac{e^{-f/3}}{3} 
           + \gamma \left(\frac{e^{-5f/6}}{6} + e^{-3f/2}\right) 
           + \gamma^2 \frac{e^{-11f/6}}{3}\ .
\end{equation}
For $N_t = 2$ and $3$, $T$ can also be derived by explicitly summing
over all possible dissociation paths. For larger $N_t$, direct
summation becomes intractable and the results following from the
general formula \eq{eq:FR_lifetime} become rather lengthy. In general,
force always affects most strongly those terms of highest order in
$\gamma$, thus for $\gamma > 1$, application of force is therefore an
efficient way to reduce average lifetime $T$. For $\gamma < 1$, $T$ is
dominated by those terms of lowest order in $\gamma$, thus here the
reduction of lifetime with increasing force is not modulated by
rebinding.

\FIG{fig:FR_lifetimeF}%
{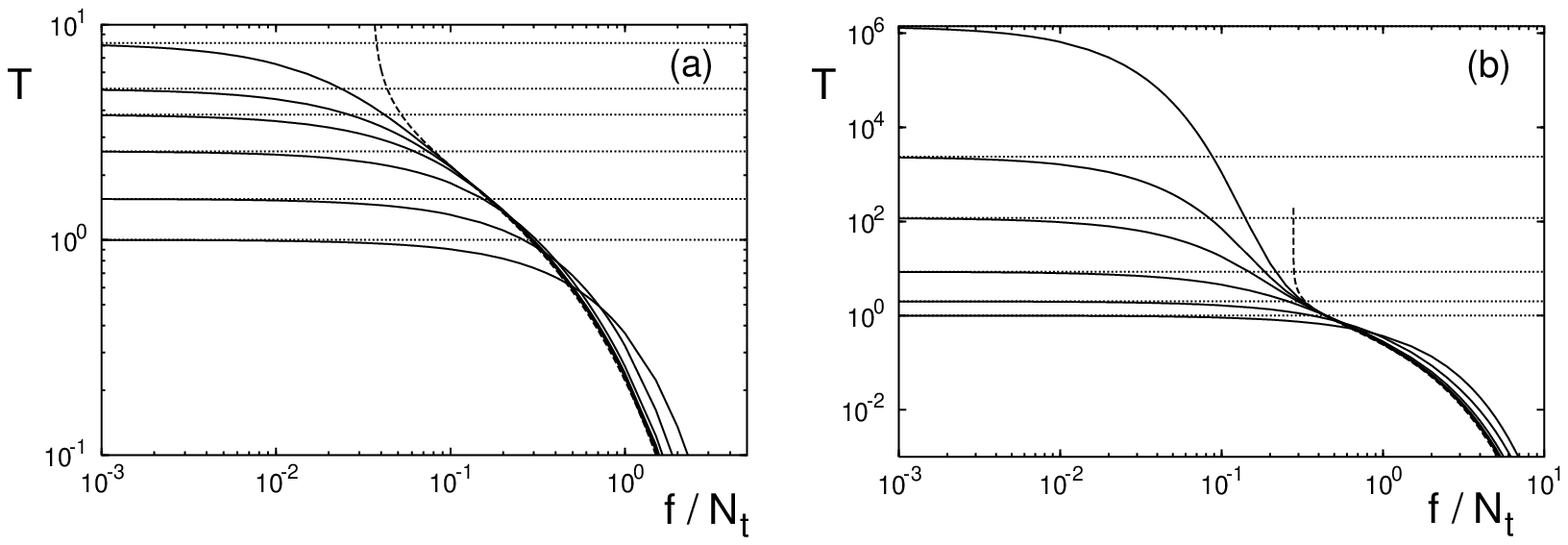}%
{Average lifetime $T$ according to \eq{eq:FR_lifetime} (solid lines)
of adhesion clusters with $N_t = 1, 2, 5, 10, 15$ and $25$ as a
function of $f/N_t$ for (a) $\gamma = 0.1$ and (b) $\gamma = 1$. The
critical forces for these rebinding rates are $f_c/N_t = 0.0355$ and
$0.278$, respectively, where the deterministic results for the
lifetimes (dashed lines) diverge.}

\fig{fig:FR_lifetimeF} shows the average lifetime of adhesion clusters of size 
$N_0 = 1, 2, 5, 10, 15$ and $25$ as a function of force-size ratio
$f/N_t$ for the rebinding constants $\gamma = 0.1$ and $\gamma =
1.0$. For small forces, $f < 1$, the average lifetime plateaus at the
value given by \eq{eq:VF_lifetime}.  For large forces, $f > N_t$, that
is, when the force on each single bond is larger than the intrinsic
force scale, the limit of vanishing rebinding applies (for $N_t = 1$
lifetime is independent of $\gamma$, compare also
\fig{fig:lifetime}). The critical forces for the given rebinding rates
are $f_c = 0.0355 N_t$ and $f_c = 0.278 N_t$. In the intermediate
force range, roughly around $f_c$, the lifetime is reduced from the
zero force to the zero rebinding limit.  This reduction is dramatic
for large clusters ($N_t \geq 10$) with appreciable rebinding ($\gamma
\geq 1$), where the lifetime is reduced by orders of
magnitude. We also show the lifetime following from the deterministic
framework, which provides a lower limit for the lifetime at large
forces, because here the largest clusters have the shortest lifetimes
for a given force size ratio $f/N_t$. Below the critical force the
deterministic lifetime is infinite and the stochastic curves approach
the plateaus \eq{eq:VF_lifetime} determined by fluctuations towards
the absorbing boundary.

\FIG{fig:FR_lifetime_NF}
{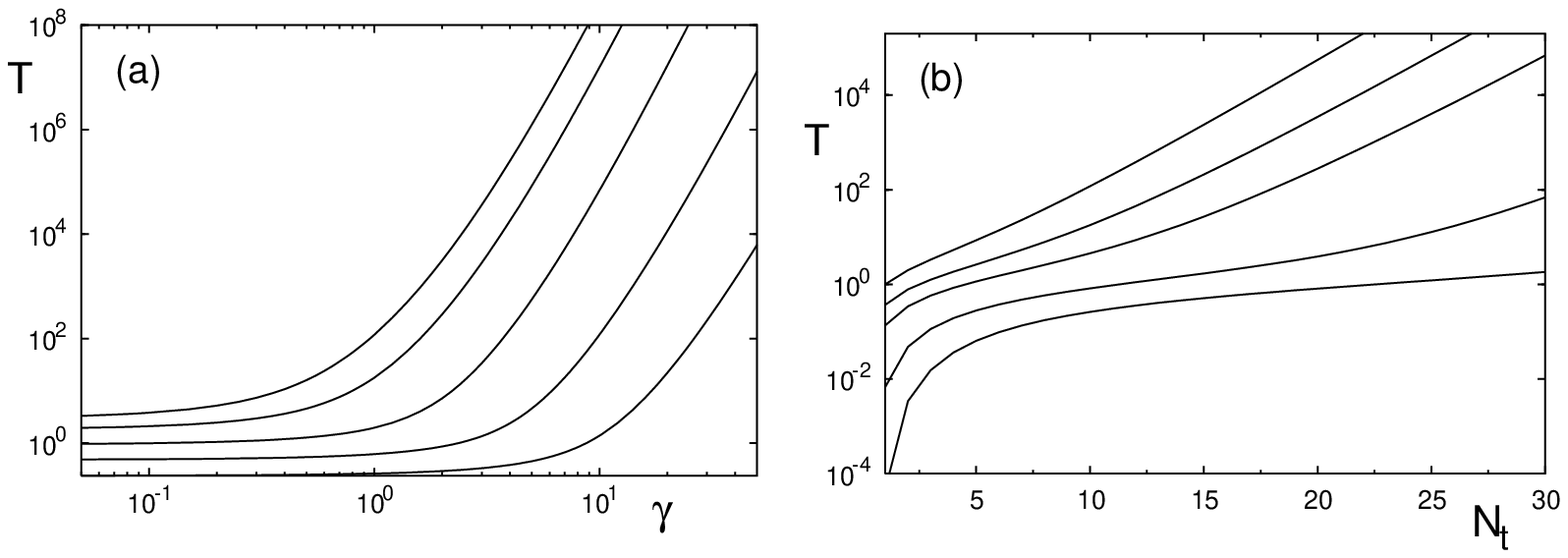}
{Average cluster lifetime $T$ (a) as function of rebinding rate
$\gamma$ for $N_t = 10$ and $f/N_t = 0, 0.1, 0.3, 0.6$ and $1$ and (b)
as function of cluster size $N_t$ for $\gamma = 1.0$ and $f = 0, 1, 2$ and
$10$.}

\fig{fig:FR_lifetime_NF}a demonstrates the influence of rebinding 
on the average lifetime at different levels of force. Here we show
average lifetime $T$ as function of $\gamma$ for $N_t = 10$
and for increasing values of force. For $f = 0$ the curves are as
depicted in \fig{fig:VF_lifetime}.  Increasing force reduces the
lifetime strongly and leads to an almost constant lifetime for
different $\gamma$ (compared the strong increase for $f = 0$). Only
when rebinding is sufficiently strong that force is smaller than the
critical force, $f < f_c$, lifetime begins to grow.  The increase
observed then is similar to that for vanishing force, only that the
absolute value of lifetime is smaller. For example, for $f = 0.6 N_t$,
the cluster grows strongly for $g \geq 5$ where the critical force is
$f_c = 0.82 N_t$; for $f = N_t$ the strong increase is observed for $g
\geq 10$, for which the critical force is $f_c = 1.15 N_t$.
A similar effect is observed for the dependence of average lifetime on
cluster size, see \fig{fig:FR_lifetime_NF}b.  At small $N_t$, cluster
lifetime grows strongly at large forces according to \eq{eq:lifetime}
due to shared loading. For larger $N_t$, lifetime grows slowly until
$N_t$ is large enough that $f_c \geq f$ is reached.  Above this size,
$T$ grows on a rate comparable to that for vanishing or small
force. For $\gamma = 0.1$, the increase of $T$ with $N_t$ is slow
throughout the shown range of $N_t$.

\section{Discussion}
\label{sec:discussion}

In this paper, we have presented a detailed analysis of the stochastic
dynamics of an adhesion cluster of size $N_t$ under shared loading $f$
and with rebinding rate $\gamma$. The corresponding master equation
has been solved exactly for several special cases. For vanishing
rebinding ($\gamma = 0$), the exact solution \eq{eq:tees20_a} could be
constructed because cluster decay is a sequence of Poisson
processes.  For vanishing force ($f = 0$), we deal with a linear
problem, which can be treated with standard techniques. In the case of
natural boundaries (that is for a reflecting boundary at $i = 0$), the
exact solution \eq{eq:VF_states} follows with the help of a generating
function. In the general case of finite force $f$ and finite rebinding
rate $\gamma$, for the case $N_t = 2$ and an absorbing boundary we
used an eigenvalue analysis to derived the exact solution
\eq{eq:two_bonds_exact}. In principle, the same method can also be
applied for a reflecting boundary or for larger clusters, but
this does not lead to simple analytical results.

For vanishing force ($f = 0$) and an absorbing boundary at $i = 0$, we
introduced the `leakage approximation' (also known as 'local thermal
equilibrium description' in the theory of protein folding), which
treats the absorbing boundary as a small perturbation to the exactly
solved case of the reflecting boundary. The resulting formulae given
in \eq{eq:LTEc} work well if average cluster lifetime $T$ is much
larger than the internal time scale $1/(1+\gamma)$ (that is for large
clusters or strong rebinding).  All other cases have been treated
with exact stochastic simulations using the Gillespie algorithm, which
for large clusters is more efficient than the eigenvalue
analysis. Moreover, the study of single simulation trajectories offers
valuable insight into the typical nature of unbinding trajectories
expected for experiments.

Once the master equation is solved, either exactly or numerically, all
quantities of interest can be calculated. In this paper, we focused on
the mean number of closed bonds as a function of time, $N(\tau)$, and
the dissociation rate for the overall cluster, $D(\tau)$. The first
moment of $D(\tau)$ then gives the mean cluster lifetime $T$. In this paper,
we derived an exact solution $T = T(N_t,f,\gamma)$ from the adjoint
master equation, see \eq{eq:FR_lifetime}. For the special cases of
vanishing rebinding and vanishing force, we also showed how the exact
formulae for $T$ can be derived via completely different routes. The
result for $T = T(N_t,f)$ from \eq{eq:lifetime} follows from the
unique dissociation path without rebinding, while the result for $T =
T(N_t,\gamma)$ from \eq{eq:VF_lifetime} can be derived with Laplace
techniques as a mean first passage time for the case of a reflecting
boundary.  In order to assess the role of fluctuations, we also
calculated the standard deviations $\sigma_N$ and $\sigma_T$ for the
distributions of the number of closed bonds and cluster lifetimes,
respectively.

A special focus of this paper was a detailed comparision between the
stochastic and determinstic treatment. Regarding mean cluster
lifetime, the deterministic treatment is rather good in the case of
vanishing rebinding, although it underestimates the plateau value for
cluster lifetime at small force. In the presence of rebinding, the
deterministic treatment fails, because it includes neither the effect
of fluctuations nor the effect of an absorbing boundary.  In
particular, the deterministic treatment does not predict finite
lifetime below the critical force $f_c$, when clusters decay due to
fluctuations towards the absorbing boundary.  Only at very large
force, when rebinding becomes irrelevant, does the deterministic
treatment work well again. Regarding the average number of closed
bonds, the deterministic model fails because it does not correctly
treat the non-linearity in the rupture rate. This effect is most
evident for small clusters and at late stage of rupture. In general,
the mean number of closed bonds in the stochastic model decay in a
smoother way than in the deterministic model, which typically shows an
abrupt decay in late stage. This abrupt decay in fact is typical for
shared loading and shows up in the stochastic model when one studies
single simulation trajectories. In this sense, the deterministic model
makes an interesting prediction which should be confirmed in
experiment, albeit not on the level of the first moment, as suggested
by the deterministic model, but rather on the level of single
trajectories, as suggested by the stochastic model.

Our results can now be used to evaluate a large range of different
experimental situations. The stochastic dynamics of adhesion clusters
under force can be quantitatively studied with many different
techniques, including atomic force microscopy, optical tweezers,
magnetic tweezers, the biomembrane force probe, flow chambers, and the
surface force apparatus. In all of these cases, by measuring cluster
lifetime $T$ and two out of the three parameters $N_t$, $f$ and
$\gamma$, the third parameter can be estimated with the help of our
exact results. In general, our exact results nicely show how mean
cluster lifetime $T$ varies with cluster size $N_t$, force $f$ and
rebinding rate $\gamma$.  For example, if the single bond lifetime was
one second ($k_0 = 1/s$), for $f = 0$ and $\gamma = 0$ a cluster
lifetime $T$ of one minute could only be achieved with $10^{26}$
bonds, because in this case, cluster lifetime scales only
logarithmically with cluster size.  However, for a rebinding rate
$\gamma = 1$ ($k_{on} = k_0$), only $N_t = 10$ bonds are necessary,
because lifetime scales strongly with rebinding, $T \sim
\gamma^{N_t-1}$. Increasing force to $f = 10$ would decrease lifetime
to $T = 0.05$ s, because $T$ is exponentially decreased by $f$. To
reach one minute again, cluster size or rebinding rate had to be
increased such that $f < f_c$. This implies $N_t > 50$ or $\gamma >
10$. It is important to note that these predictions are based on the
assumption of rigid force transducers. In many experimental situations
of interest, the force transducer will be subject to elastic
deformations or even to viscous relaxation processes, like for example
when pulling on cells \cite{c:beno00}. In order to focus on generic
aspects of adhesion clusters, here we only studied the minimal model
for stochastic dynamics under force.

Our results can also be applied to experiments in cell adhesion.  For
example, the biomembrane force probe with linear loading has recently
been used to study the decay of $\alpha_{\nu} \beta_3$-integrin
clusters induced on the surface of endothelial cells
\cite{c:prec02,uss:erdm04b}. If one makes sure that the clusters do
not actively grow during the time of dissociation, similar experiments
could now be done also for constant loading. Because in these kinds of
experiments the exact cluster size is usually unknown, one had to
convolute our results with a Poisson distribution for an estimated
average number of bonds \cite{c:ches98,c:zhu00}. Recently, our result
for the average cluster lifetime of two bonds under shared force and
with rebinding, \eq{eq:T_two_bonds_exact}, has been applied to the
analysis of flow chamber data on leukocyte tethering through
L-selectin \cite{uss:schw04a}. Since in this case force can be
calculated as a function of shear flow, our formula can
be used to the estimate rebinding rate, which in this case turns out
to be surprisingly large. This in turn explains why dissociation
dynamics in L-selectin mediated leukocyte tethering appears to be
first order: for large rebinding, the leakage approximation is rather
good, and decay is exponential.

Our results can not be directly applied to adhesion clusters which
compensate for force-induced decay by active growth, as it has been
found experimentally for focal adhesions \cite{uss:rive01}.  Yet there
are also interesting lessons for focal adhesions which can be learned
from our model. For example, our stochastic analysis confirms the
prediction from the deterministic stability analysis that cluster
stability changes strongly around the critial value $f_c$ (although
small clusters tend to decay also at smaller force due to fluctuations
towards the absorbing boundary). It is interesting to note that recent
experiments measuring internally generated force at single focal
adhesions suggest that $f/N_t$, the most important scaling variable of
our analysis, is roughly constant for different cell types
\cite{uss:bala01,c:tan03}. It is therefore tempting to speculate that
focal adhesions (or subsets of focal adhesions) are regulated to be
loaded close to the critical value $f_c / N_t = \rm plog (\gamma / e)$
from \eq{eq:fcrit}. In this way, cells could quickly increase force on
single bonds by small changes in actomyosin contractility. Large force
on single closed bonds in turn might trigger certain signaling events
in focal adhesions, possibly by mechanically opening up certain
signaling domains \cite{c:isra01}. Our speculation provides a simple
way to estimate the rebinding rate, which is very hard to measure
experimentally. Using compliant substrates, it has been found that
focal adhesions are characterized by a stress constant $\sim 5.5$
nN/$\mu m^2$ \cite{uss:bala01,c:tan03}. We do not know which of the
many different proteins in focal adhesions defines the weak link which
most likely ruptures under force, but we expect that it will have a
similar area density as the integrin receptors, which are expected to
have a typical distance between 10 and 30 nm, corresponding to $10^4$
and $10^3$ molecules per $\mu m^2$, respectively. To obtain a lower
estimate for $\gamma$, we therefore use $F_c = 5.5$ nN and $N_t =
10^4$. For activated $\alpha_5 \beta_1$-integrin binding to
fibronectin, recent single molecule experiments obtained for the
molecular parameter values $k_0 = 0.012$ Hz and $F_b = 9$ pN
\cite{c:li03}. Therefore the rebinding rate can be estimated to be at
least $\gamma = 0.2$, that is $k_{on} = 0.002$ Hz in dimensional
units. Based on future experimental input, it would be interesting to
extend our model of passive decay to active processes resulting in
cluster growth under force.

Finally we want to comment that our model might also be applied to
situations in materials science which are not directly related to
biomolecular receptor-ligand pairs. One example is sliding friction,
which recently has been modeled as dynamic formation and rupture of
bonds under force \cite{c:fili04}.  In general, we expect that many
more cohesion phenomena in materials can be successfully modeled as
dynamic interplay between rupture and rebinding.
 
\textbf{Acknowledgments:} This work was supported by the German
Science Foundation through the Emmy Noether Program.

%\bibliographystyle{apsrev}
%\bibliography{b,c,r,uss}

\end{document}